\documentclass[useAMS]{mn2e}
\usepackage{graphicx}

                              \def\apj{ApJ}          \def\apjl{ApJ}                              \def\apss{Ap\&SS}          \def\aap{A\&A}                                                                                                              \def\mnras{MNRAS}                                                                                                              \def\pasp{PASP}                    \def\qjras{QJRAS}

\def\degr{\hbox{$^\circ$}}

\def\arcsec{\hbox{$^{\prime\prime}$}}

\title[Radiative transfer modelling of  IRDCs]
{Modelling {\it Herschel} observations of infrared-dark clouds in the Hi-GAL survey}
\author[D. Stamatellos et al.]
{D. Stamatellos$^{\!  1}$\thanks{E-mail:D.Stamatellos@astro.cf.ac.uk}\thanks{Herschel is an ESA space observatory with science instruments provided by European-led Principal Investigator consortia and with important participation from NASA.} , M.~J. Griffin$^{\!  1}$, J.~M. Kirk$^{\!  1}$,  S. Molinari$^{\!  3}$ , 
B. Sibthorpe$^{\!  2}$,
\newauthor {D. Ward-Thompson$^{\!  1}$, A.~P. Whitworth$^{\!  1}$, 
L.~A. Wilcock$^{\!  1}$}\\ 
$^1$ School of Physics \& Astronomy,Cardiff University, Cardiff, CF24 3AA, 
Wales, UK\\
$^2$ UK Astronomy Technology Centre, Royal Observatory, Blackford Hill, 
Edinburgh EH9 3HJ, UK\\
$^3$ INAF-IFSI, Fosso del Cavaliere 100, 00133 Roma, Italy}
\begin{document}

\date{Accepted 2010 January 15. Received 2010 January 2; in original form 
2010 January 1}

\pagerange{\pageref{firstpage}--\pageref{lastpage}} \pubyear{2010}

\maketitle

\label{firstpage}

\begin{abstract}
We demonstrate the use of the 3D Monte Carlo  radiative transfer code {\sc phaethon} to model infrared-dark clouds (IRDCs) that are externally illuminated by the interstellar radiation field (ISRF). These clouds are believed to be the earliest observed phase of high-mass star formation, and may be the high-mass equivalent of lower-mass prestellar cores. We model three different cases as examples of the use of the code, in which we vary the mass, density, radius, morphology and internal velocity field of the IRDC. We show the predicted output of the models at different wavelengths chosen to match the observing wavebands of {\it Herschel} and {\it Spitzer}. For the wavebands of the long-wavelength SPIRE photometer on {\it Herschel}, we also pass the model output through the SPIRE simulator to generate  output images that are as close as possible to  the ones that would be seen using SPIRE. We then analyse the images as if they were real observations, and compare the results of this analysis with the results of the radiative transfer models. We find that detailed radiative transfer modelling is necessary to accurately determine the physical parameters of IRDCs (e.g. dust temperature, density profile). This method is applied to study G29.55+00.18, an IRDC observed by the {\it Herschel} Infrared Galactic Plane survey (Hi-GAL), and in the future it will be used to model a larger sample of IRDCs from the same survey.
\end{abstract}


\section{Introduction}

High-mass stars evolve somewhat differently from low-mass stars. There appear to be (at least) four main evolutionary stages of high-mass star formation (e.g. Zinnecker \& Yorke 2007): infra-red-dark clouds; hot molecular cores; compact HII regions (including hyper- and ultra-compact HII regions); and classical HII regions.

The earliest observed phase of high-mass star formation is the infra-red-dark cloud (IRDC) stage. IRDCs are very dense, massive interstellar clouds in which high-mass stars are believed to form. They were first identified in infra-red surveys, by the {\it Infrared Space Observatory} ({ISO}; Perault et al. 1996)  and the {\it Midcourse Space Experiment} ({MSX}; Egan et al. 1998), as denser, higher-mass and more distant analogues to prestellar cores (Ward-Thompson et al. 1994, 2007), from which lower-mass stars form (e.g. Solar-type stars).  In fact,  IRDCs are so dense that they are optically thick even at infra-red wavelengths ($\sim$8--24$\mu$m). In many of these clouds there is little or no detectable star formation (e.g. Parsons et al. 2009). 

The mass of an IRDC can be hundreds, or even thousands of solar masses (e.g. Rathborne et al. 2009). Often, dense, dark cores can be seen within each cloud. These are known as infrared-dark cores, and can have masses of up to $\sim$100~M$_\odot$, and are typically only 0.1~pc or less in radius. Hence they have densities of up to $\sim$10$^{12}$ hydrogen molecules~m$^{-3}$ (e.g. Egan et al. 1998; Carey et al. 1998, 2000). These are the most likely candidate sites of high-mass star formation.

The {\it Herschel Space Observatory} was launched on 2009 May 14, and carries a  3.5-metre diameter telescope, and three scientific instruments designed to carry out imaging and spectroscopy (Pilbratt et al. 2008, 2010). The Spectral and Photometric Imaging Receiver (SPIRE\footnote{SPIRE has been developed by a consortium of institutes led by Cardiff University (UK) and including Univ. Lethbridge (Canada); NAOC (China); CEA, LAM (France); IFSI, Univ. Padua (Italy); IAC (Spain); Stockholm Observatory (Sweden); Imperial College London, RAL, UCL-MSSL, UKATC, Univ. Sussex (UK); and Caltech, JPL, NHSC, Univ. Colorado (USA). This development has been supported by national funding agencies: CSA (Canada); NAOC (China); CEA, CNES, CNRS (France); ASI (Italy); MCINN (Spain); SNSB (Sweden); STFC (UK); and NASA (USA).}) is the long-wavelength camera on board {\it Herschel} (Griffin et al. 2009, 2010). The Hi-GAL survey is a {\it Herschel} Open Time key project that is mapping the inner Galaxy  in five bands between 60 and 500~$\mu$m (Molinari et al. 2010a,b).  Hi-GAL will map the area common to the {\it Spitzer} GLIMPSE \& MIPSGAL (Rieke et al. 2004) and MSX (Price et al. 2001) Galactic plane surveys, but at FIR and submm wavelengths. Hence, the IRDCs will be seen in emission, in contrast with the previous surveys. One of the science goals of Hi-GAL is to determine the physical properties of IRDCs and hence constrain the initial conditions of high-mass star formation.

In this paper we demonstrate the use of the 3D Monte Carlo radiative transfer code {\sc phaethon} in converting two-dimensional imaging data of IRDCs from the Hi-GAL survey, into the physical parameters of the clouds being studied, with particular emphasis on the data from SPIRE. In Section 2 we describe the radiative transfer code and the specific input parameters used in this study. In Section 3 we present radiative transfer models of three IRDCS with different morphologies (spherical, flattened, and turbulent). In Section 4 we use the SPIRE photometer simulator to produce simulated observations of the IRDC models, and in Section 5 we compare the radiative transfer models with the simulated observations. In Section 6, we apply this method to determine the properties of
G29.55+00.18, an IRDC from the Hi-GAL survey. Finally, in Section 7 we summarize  our results.

\section{Numerical method: Monte Carlo radiative tranfer }

\subsection{Code overview}

The radiative transfer calculations are performed using {\sc phaethon}\footnote{http://www.astro.cf.ac.uk/pub/Dimitrios.Stamatellos/phaethon}, a 3D Monte Carlo radiative transfer code (Stamatellos 2003; Stamatellos \& Whitworth 2003, 2005; Stamatellos et al. 2004, 2005, 2007). The code uses a large number of monochromatic luminosity packets to represent the radiation sources in the system (stars and/or the ambient radiation field). The luminosity packets are injected into the system and  interact (are absorbed, re-emitted, scattered) stochastically with it.  If a luminosity packet is absorbed its energy is added to the local region and raises the local temperature. To ensure radiative equilibrium  this packet is re-emitted immediately with a new frequency chosen from the difference between the local cell emissivity before and after the absorption of the packet (Bjorkman \& Wood 2001; Baes et al. 2005).

The input parameters of the code are: (i) the density profile of the system, (ii) the dust properties (e.g. opacity), and (iii) the radiation sources present (e.g. ambient radiation field and/or stars). The code calculates the dust temperature, and produces observables: (i) the SED of the system at different viewing angles, and (ii) synthetic images of the system  at different viewing angles and at different wavelengths.

\subsection{External radiation field and dust properties}

A starless IRDC is assumed, hence the cloud is only heated by the 
ambient radiation field. For the ambient radiation field  we adopt a 
revised version of the Black  (1994) interstellar radiation field. 
This consists of radiation from giant stars and dwarfs, thermal emission 
from dust grains, cosmic background radiation, and mid-infrared emission 
from transiently heated small PAH grains (Andr\'e et al. 2003). This 
radiation field represents well the interstellar radiation field in the 
solar neighbourhood but it is probably an underestimate of the ambient radiation field  in the Galactic Plane. An enhanced ambient radiation field could result in higher dust temperatures at the surface of the clouds but the temperatures in the central cloud regions are not expected to be significantly higher (only a few degrees). The luminosity packets representing the ambient radiation field  (typically a few $10^{10}$ packets) are injected from the outside of the cloud with injection  points and injection directions chosen to mimic an isotropic radiation field incident on the cloud (Stamatellos et al. 2004, 2007). 

The dust grains in dense clouds are expected to coagulate and accrete ice mantles, thus we use the Ossenkopf \& Henning (1994) opacities for the standard Mathis, Rumpl \& Nordsieck (1977) interstellar grain mixture (53\% silicate \& 47\% graphite),  with grains that have coagulated and accreted thin ice mantles over a period of $10^5$ years at densities $10^6 {\rm cm}^{-3}$. We assume a gas-to-dust mass ratio of 100.

\section{Radiative transfer models of IRDCs}

We model 3 different cloud geometries: (i) a spherical cloud;
(ii) a flattened cloud; and (iii) a turbulent cloud with substructure.
We also vary the mass, density and radius. The parameter values are chosen so as to illustrate some of the variety of parameter values that the code can handle. The parameters of each of the three cases are listed in Table~1. 

\begin{table}
\centering
\caption{IRDC parameters used in the radiative transfer models.
$n_{\rm c}$: cloud central density; 
$r_{\rm 0}$:  cloud flattening radius;
$R_{\rm cloud}$: cloud extent; $M_{\rm cloud}$: cloud mass; $d_{\rm cloud}$: 
cloud distance. }
\label{tab1:params}
\renewcommand{\footnoterule}{}  \centering   
\begin{tabular}{lccc}
\multicolumn{4}{c}{}\\
&{\bf Spherical }&{\bf Flattened }&{\bf Turbulent }\\
\hline 
$n_{\rm c}$ (cm$^{-3}$)& $5 \times10^6$ & $10^7$ & $5\times10^6$\\
$r_{\rm 0}$ (pc)  & 0.08& 0.04 &0.08  \\
$R_{\rm cloud}$ (pc)& 1 & 1 &1.5 \\
$M_{\rm cloud}$  (M$_{\odot}$) & 1300 & 510 &27200 \\
$d_{\rm cloud}$ (kpc)& 2 & 2 & 2\\
\hline
\end{tabular}
\end{table}

\subsection{Spherical IRDC}

For the spherical cloud we adopt a Plummer-like  density profile (Plummer 
1915),

\begin{equation}
n(r) = n_{\rm c}\,\frac{1}{\left[1+\left(\frac{r}{r_0} \right)^2\right]^2} \,,
\end{equation} 

\noindent
where $n_{\rm c}$ is the density at the centre of the cloud, and 
$r_0$ is the extent of the region in which the density is 
approximately uniform. This density profile is consistent with observations 
of low-mass cores in nearby star forming clouds, such as L1544, L63 and L43
(e.g. Kirk et al. 2005, 2006, 2007), and has been used to model such clouds (Whitworth \& Ward-Thompson 2001). The values of these parameters are given in Table~1. The mass of the cloud is 1300~M$_{\sun}$.

\begin{figure}
\centerline{
\includegraphics[width=6cm,angle=-90]{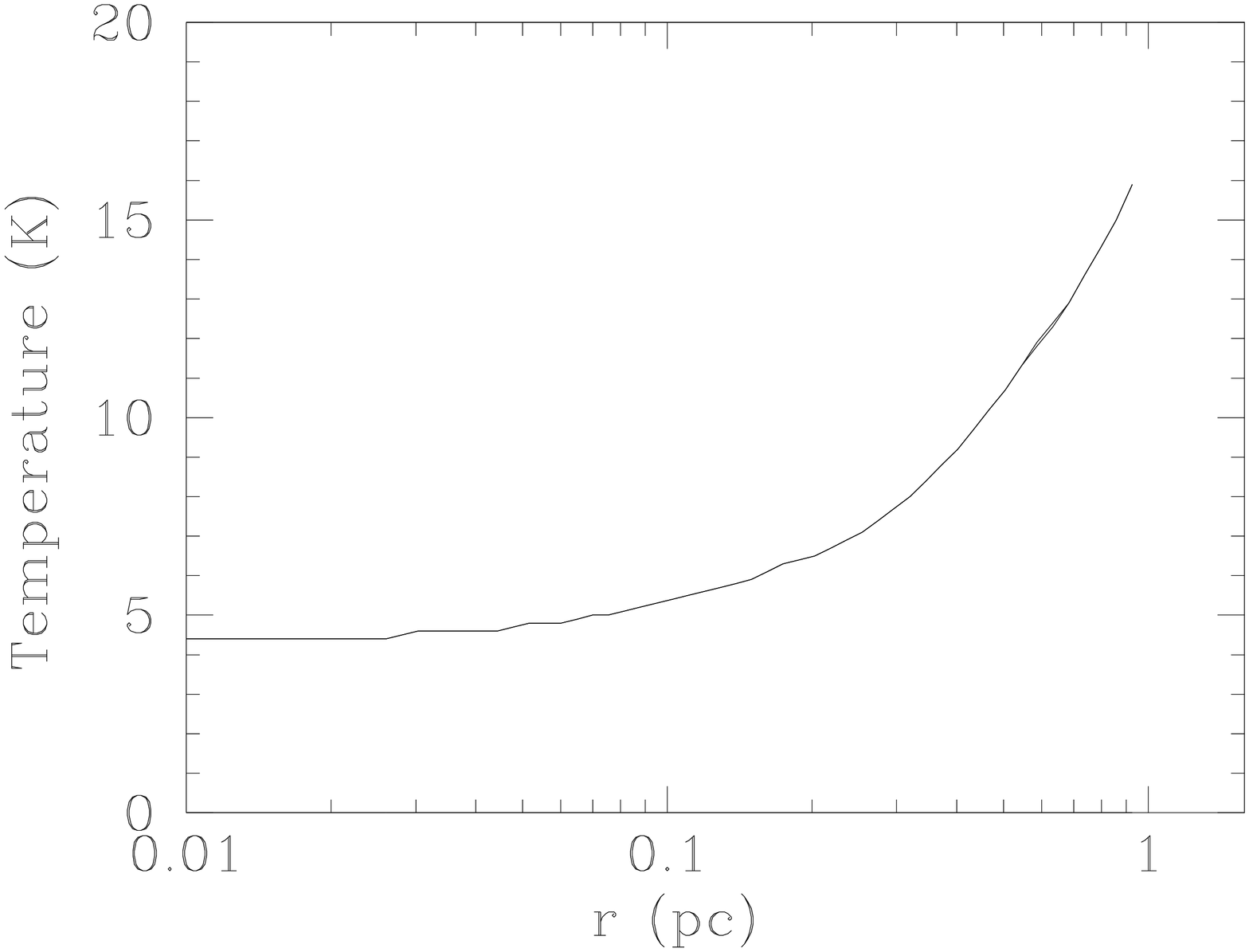}}
\caption{Dust temperature profile of a spherical IRDC.}
\label{fig1:temp}
\centerline{
\includegraphics[width=6cm,angle=-90]{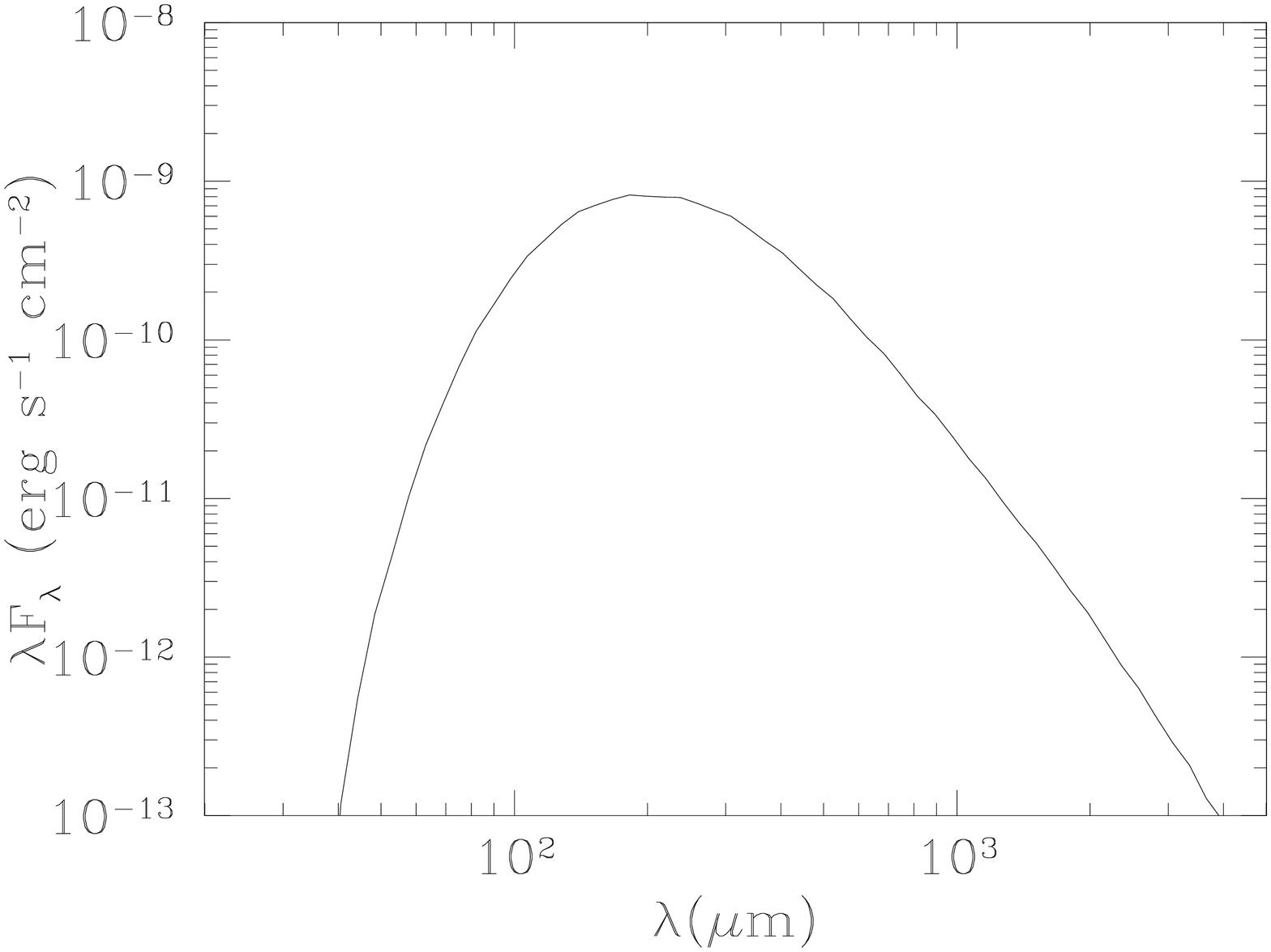}}
\caption{SED of a spherical cloud at a distance of 2~kpc.}
\label{fig2:spec}
\end{figure}

Figure~1 shows that the temperature in the cloud
drops from $\sim 16$~K at the edge of the cloud to $\sim 4$~K at the centre 
of the cloud. The central dust temperature is $2-3$~K lower than the typical temperatures in the central regions of low-mass (i.e. a 
few solar masses) clouds ($\sim 7$~K; e.g. Stamatellos \& Whitworth 2003; 
Stamatellos et al. 2004). This is due to the much higher mass
and extinction of the IRDC.  Heating from cosmic rays, which is not included in our models, may increase the temperature in the centre of the cloud to $\sim 5$~K. Additionally, a stronger external radiation field in the Galactic plane could also possibly increase the temperature by a few degrees).

Figure~2 shows that the SED of the cloud peaks at around 200~\micron, 
consistent with the low temperatures in the cloud. Because the temperature 
gradient in the IRDC is larger than the temperature gradient in a low-mass 
core, the SED is broader.

\subsection{Flattened IRDC}
\label{sec:flat}

\begin{figure}
\centerline{
\includegraphics[width=6cm,angle=-90]{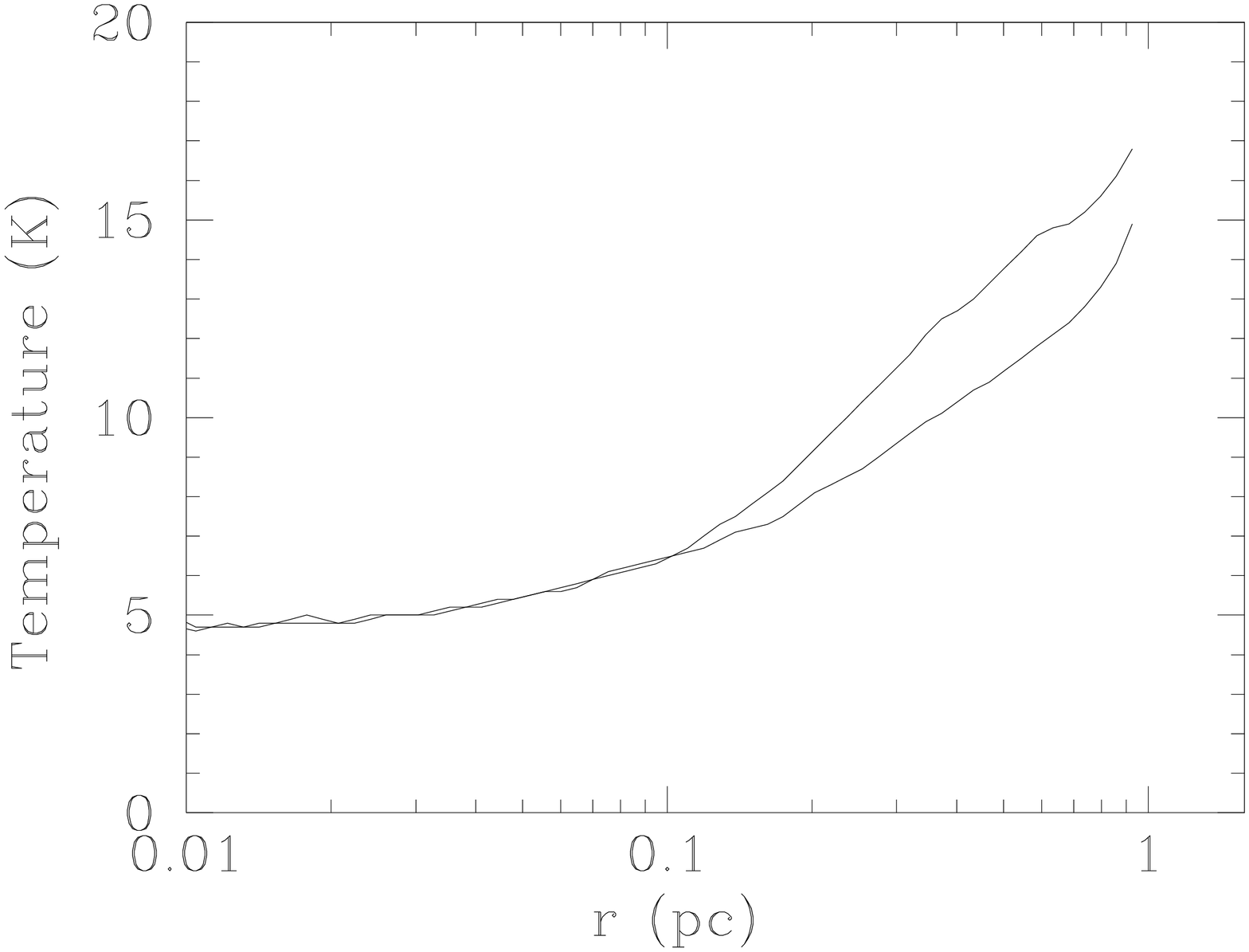}}
\caption{Dust temperature profile of a flattened cloud at two different 
directions  in the cloud. The bottom line corresponds to the midplane 
of the flattened structure, whereas the upper line corresponds to the 
direction perpendicular to the midplane.}
\label{fig3:temp}
\centerline{
\includegraphics[width=6cm,angle=-90]{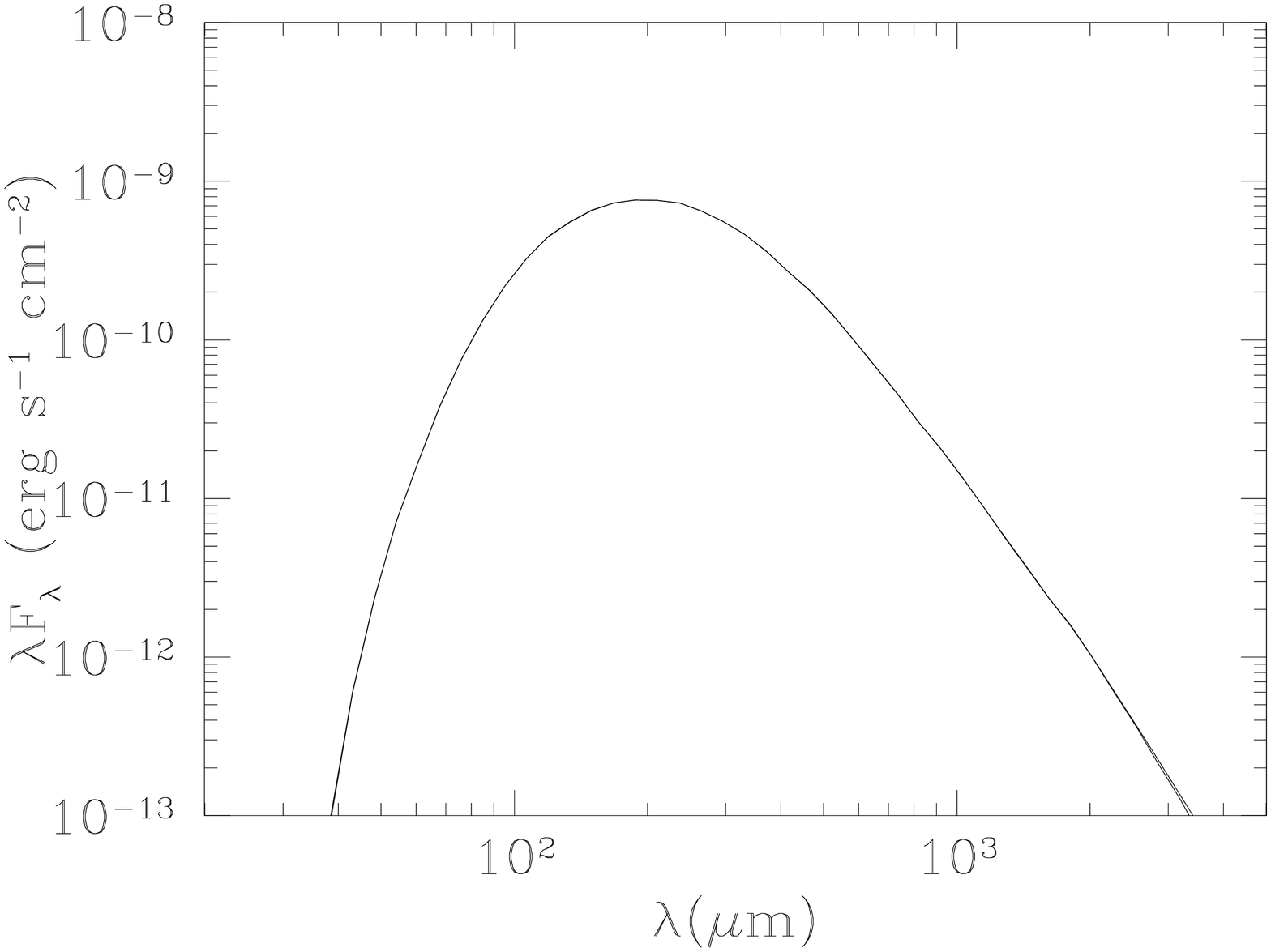}}
\caption{SED of a flattened cloud at a distance of 2~kpc. }
\label{fig4:sed}
\end{figure}

\begin{figure*}
\centerline{
\includegraphics[width=4.5cm,angle=-90]{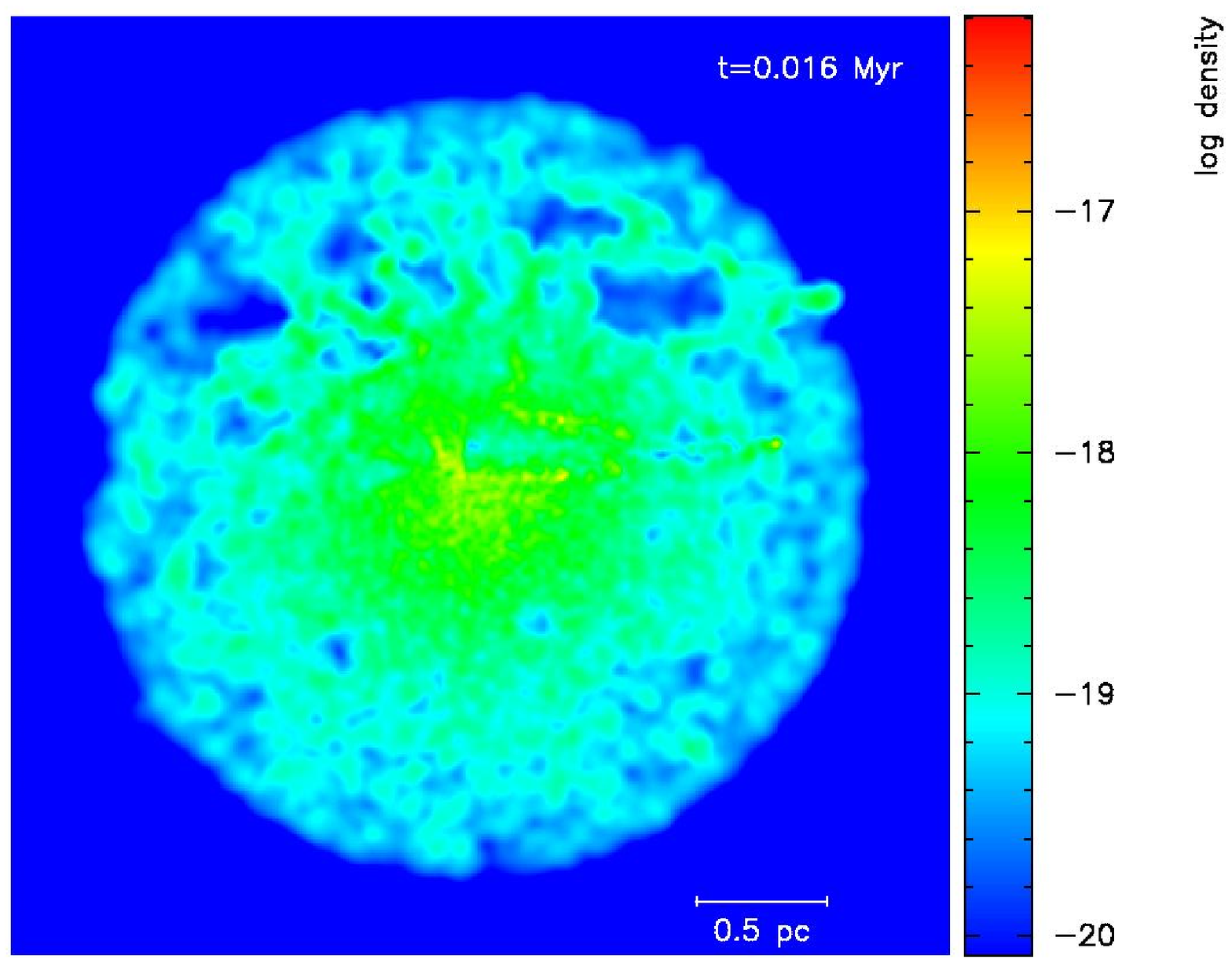}
\includegraphics[width=4.5cm,angle=-90]{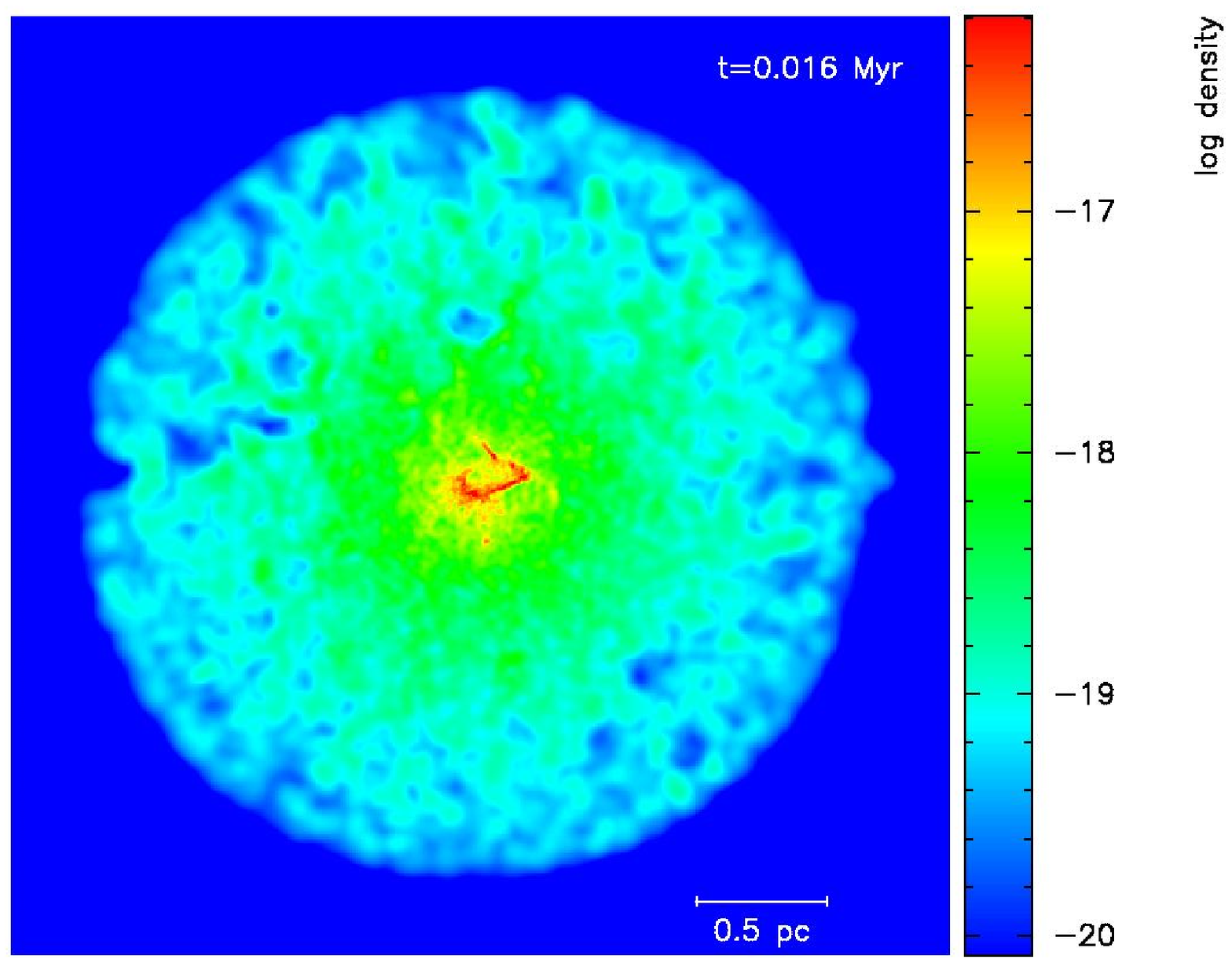}
\includegraphics[width=4.5cm,angle=-90]{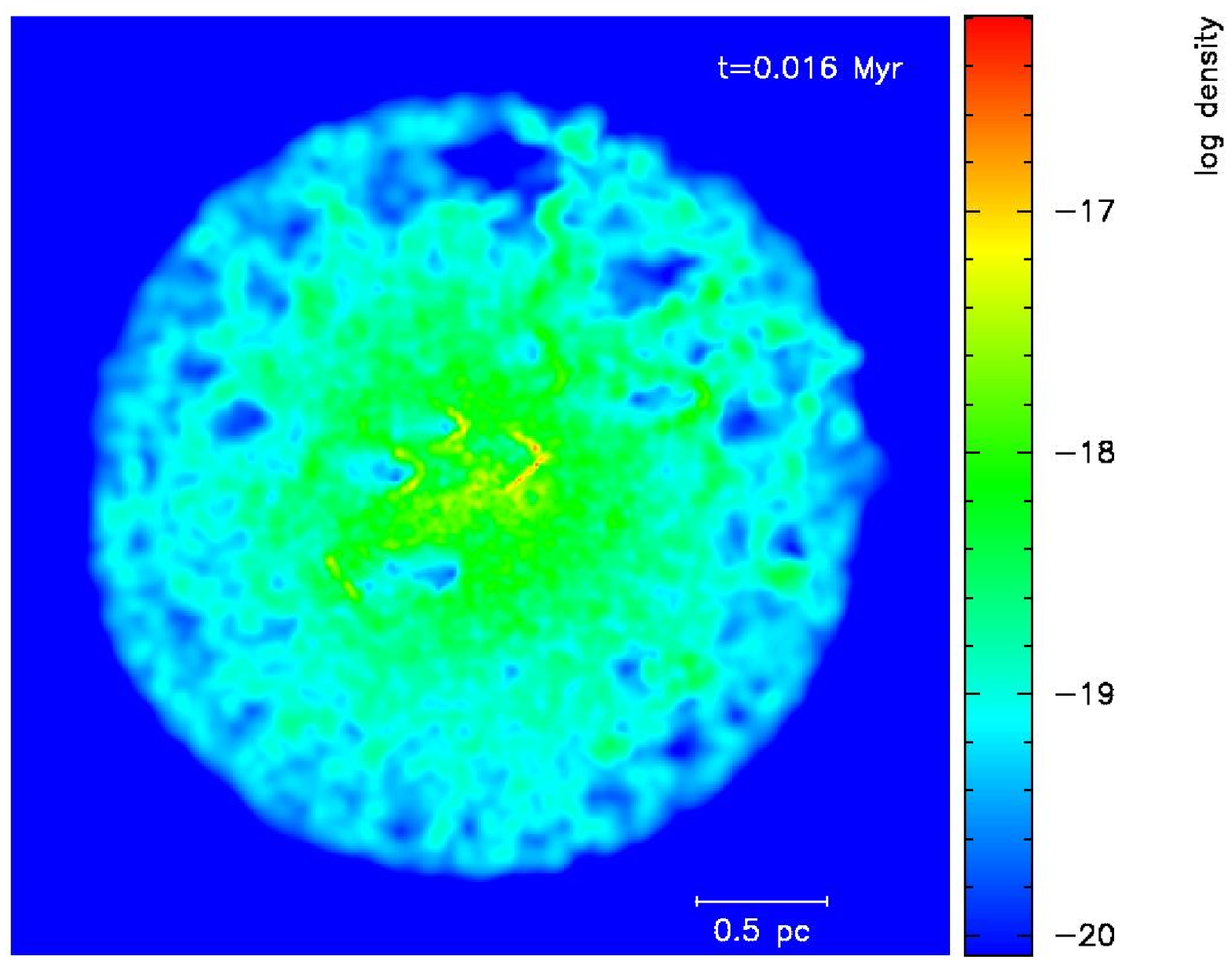}}
\caption{Density  on the $y=-0.2$ pc (left), $y=0$ pc (centre) 
and $y=+0.2$ pc (right)  planes for a turbulent IRDC. The density is given 
in units of g cm$^{-3}$.}
\label{fig5:dens}
\end{figure*}

For the flattened cloud we use a density profile of the form
\begin{equation}
\label{eq:flatdens}
n(r,\theta) = n_{\rm c}\,\frac{1 + A \left( \frac{r}{r_0} \right)^2 
{\rm sin}^p(\theta) }{ \left[ 1 + \left( \frac{r}{r_0} \right)^2 
\right]^2 } \,,
\end{equation}
where $n_{\rm c}$ is the density at the centre of the cloud, and 
$r_0$ is the extent of the region in which the density is 
approximately uniform  (see Stamatellos et al. 2004). The values of 
these parameters are given in Table~1. The parameter 
$A$ determines the equatorial-to-polar optical 
depth ratio $e\,$, i.e. the maximum optical depth from the centre 
to the surface of the cloud (which occurs at $\theta = 
90\degr$), divided by the minimum optical depth from the centre 
to the surface of the cloud (which occurs at $\theta = 0\degr$ 
and $\theta = 180\degr$). The parameter $p$ determines how rapidly 
the optical depth from the centre to the surface rises with 
increasing $\theta$, i.e. going from the north pole at $\theta 
= 0\degr$ to the equator at $\theta = 90\degr$. In this model we assume 
$e=1.03$, i.e. a slightly flattened cloud, and $p=4$. This geometry may 
be more realistic than the spherical cloud presented in the previous section.
The mass of the cloud is 510~M$_{\sun}$ (see Table~1).

Figure~3 shows that the temperature profile in the cloud
is similar to the case of the spherical cloud; the temperature drops 
from 18 K at the edge to 5 K in the centre of the cloud. However, due 
to the flattened cloud geometry the cloud  `equator' is colder than 
the cloud `poles'.

Figure~4 shows that the SED of the cloud is also similar to that of the spherical cloud. There is no dependence on the viewing angle 
despite the fact that the optical depth to the centre of the cloud becomes 
$\tau_{\rm cloud}\sim 1$ at $\sim 200~\micron$. This is because the 
temperature varies with the direction in the cloud only in the outer 
region of the cloud, which is optically thin even at wavelengths up to
$\sim 200~\micron$.

\subsection{Turbulent IRDC}

We finally examine a more realistic cloud geometry, i.e. a turbulent cloud. To produce this cloud we start off with a spherical cloud having a density profile 
\begin{equation}
n(r) = n_{\rm c}\,\frac{1}{1+\left(\frac{r}{r_0} \right)^2} \,,
\end{equation}
where $n_{\rm c}$ is the density at the centre of the cloud, and 
$r_0$ is the extent of the region in which the density is  approximately uniform. Note that this profile is less steep than the profile used for the spherical IRDC, hence the higher mass of this cloud.

The values of the parameters are again given in Table~1. The mass of the cloud is 27200 M$_{\sun}$. A large turbulent velocity field  ($\alpha=0.5$) is imposed to the cloud (e.g. Goodwin et al. 2004), and the cloud is evolved using the SPH code DRAGON, for enough time to produce substructure, i.e. until cores form in the cloud (e.g. Stamatellos et al. 2007).  DRAGON is a gravitational hydrodynamics code which invokes a
large number of particles to represent a physical system. It uses an  octal tree (to compute gravity and find neighbours), adaptive smoothing lengths, multiple particle timesteps, and a second-order Runge-Kutta integration scheme. The resulting  cloud is shown in Figure~5, in which we plot its density on 3 planes ($y=-0.2$ pc, left, $y=0$ pc, centre and $y=+0.2$ pc, right). 

We perform a radiative transfer simulation for this cloud using the method of Stamatellos \& Whitworth (2005). The calculated  dust temperature  is presented in Figure~6. The dust temperature is similar to the temperature calculated in the previous cases. However, the temperature distribution at a particular distance from the centre of the cloud is broader due to the clumpiness of the cloud. Figure~7 shows that the SED of the cloud peaks at longer wavelengths than in the previous two cases,  as the cloud is more massive  and consequently it is cooler.

\begin{figure}
\centerline{
\includegraphics[width=6.cm,angle=-90]{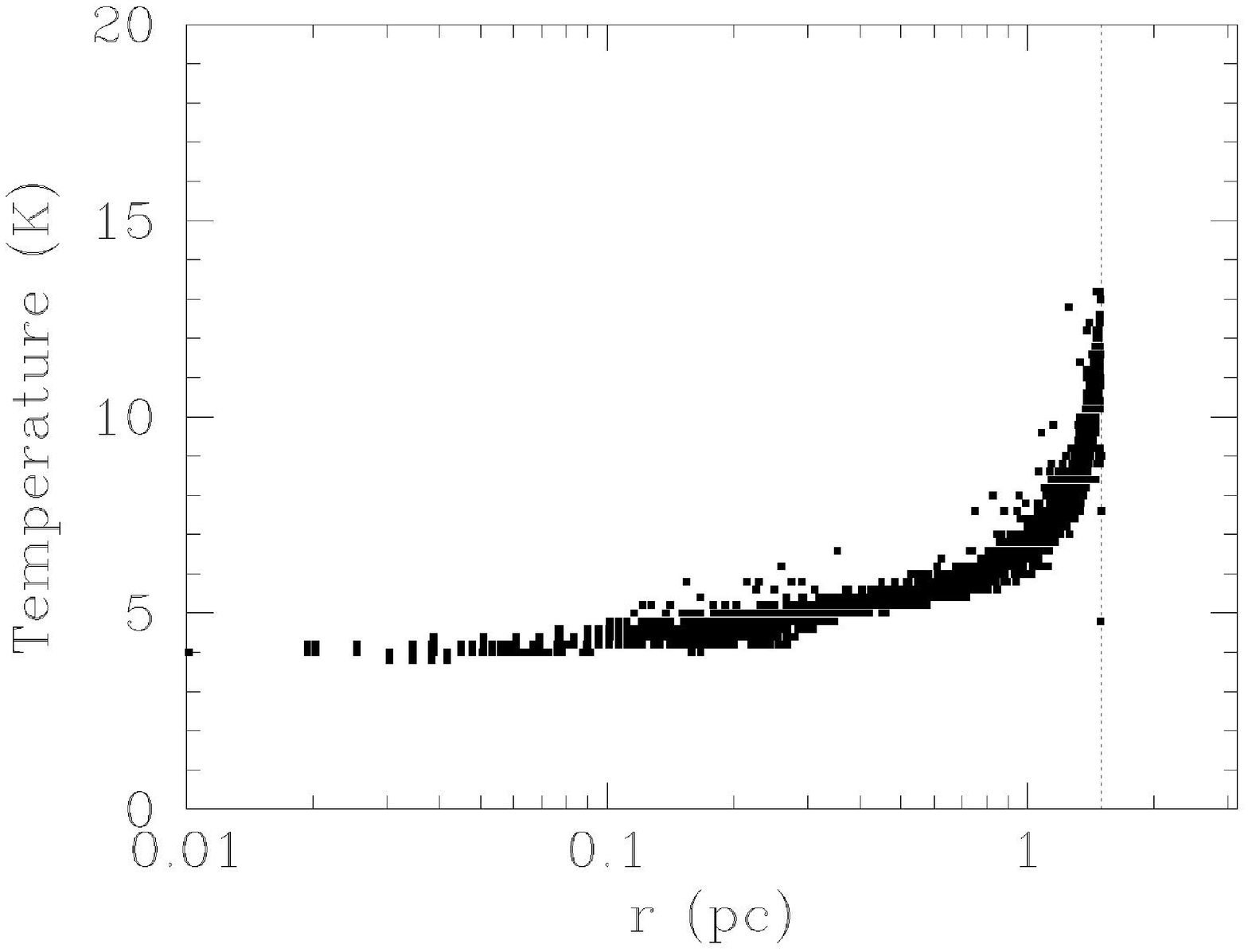}}
\caption{Dust temperature of a turbulent IRDC. The points correspond to different locations in the cloud.}
\label{fig:msx3d.temp}
\centerline{
\includegraphics[width=6cm,angle=-90]{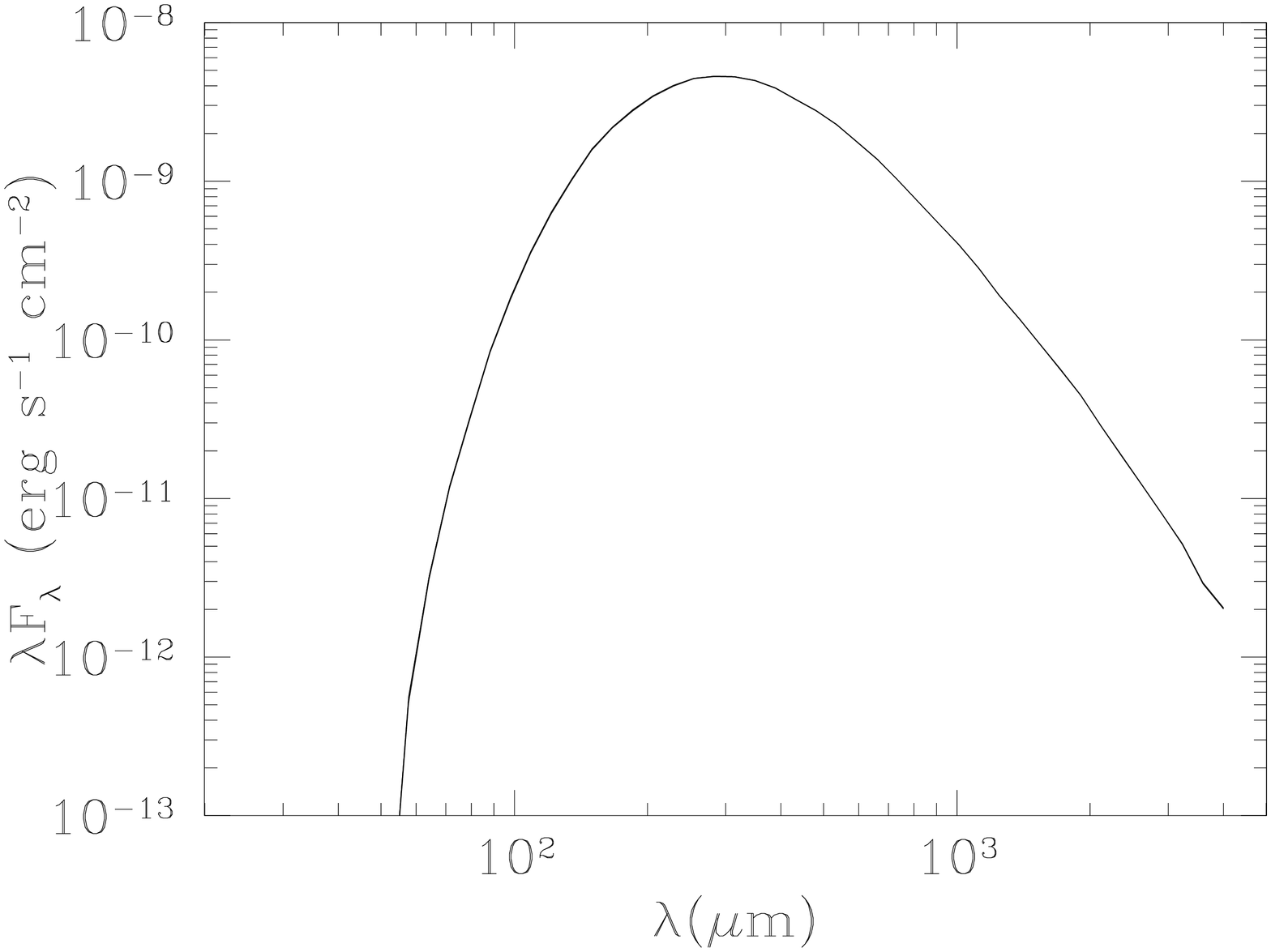}}
\caption{SED of a turbulent IRDC at a distance of 2~kpc.}
\label{fig:msx3d.spec}
\end{figure}

\section{SPIRE simulated observations}
\label{sec:spire}

Simulated observations of the modelled IRDCs are performed, to assess the impact of instrument systematics. The simulations are produced using V2.31b of the SPIRE photometer simulator (Sibthorpe, Chanial and Griffin 2009).  This software simulates both the SPIRE instrument, and the {\it Herschel} observing modes. The simulations are performed in a configuration which matches the Hi-GAL observations.  

The fast scanning speed (60\arcsec per second) option is used (see the SPIRE Observers' Manual\footnote {http://herschel.esac.esa.int/Documentation.shtml}), and the observations are made in the SPIRE parallel mode, to match the Hi-GAL mapping mode. These observations are performed by continuously scanning the telescope backwards and forwards in a raster pattern while the instruments continuously take data.  

For Hi-GAL two maps are obtained by scanning first along one axis of the instrument array, and then scanning along the perpendicular axis of the array.  This provides redundancy in the data, allowing for the use of maximum likelihood techniques to be used in the subsequent data reduction (Sibthorpe et al. 2008).

Realistic noise with a 1/\emph{f} spectrum is simulated with each of the SPIRE bolometers using independent noise parameters based on pre-flight test data. Thermal drifts are not included in these simulations as the information is not available to model these accurately.  However, the {\it Herschel}/SPIRE pipeline will automatically remove these drifts, making their omission from these simulations insignificant.

The simulated data are calibrated and reduced from time-line data to maps using a simple map-maker.  Maps at the three SPIRE bands, 250, 350 and 500\,$\mu$m are obtained, having full width half maximum beams sizes 
of 18, 25 and 36 arcseconds respectively.

\section{Comparison of IRDC  models to the simulated observations}

\subsection{Maps}

\begin{figure*}
	\includegraphics[width=\textwidth]{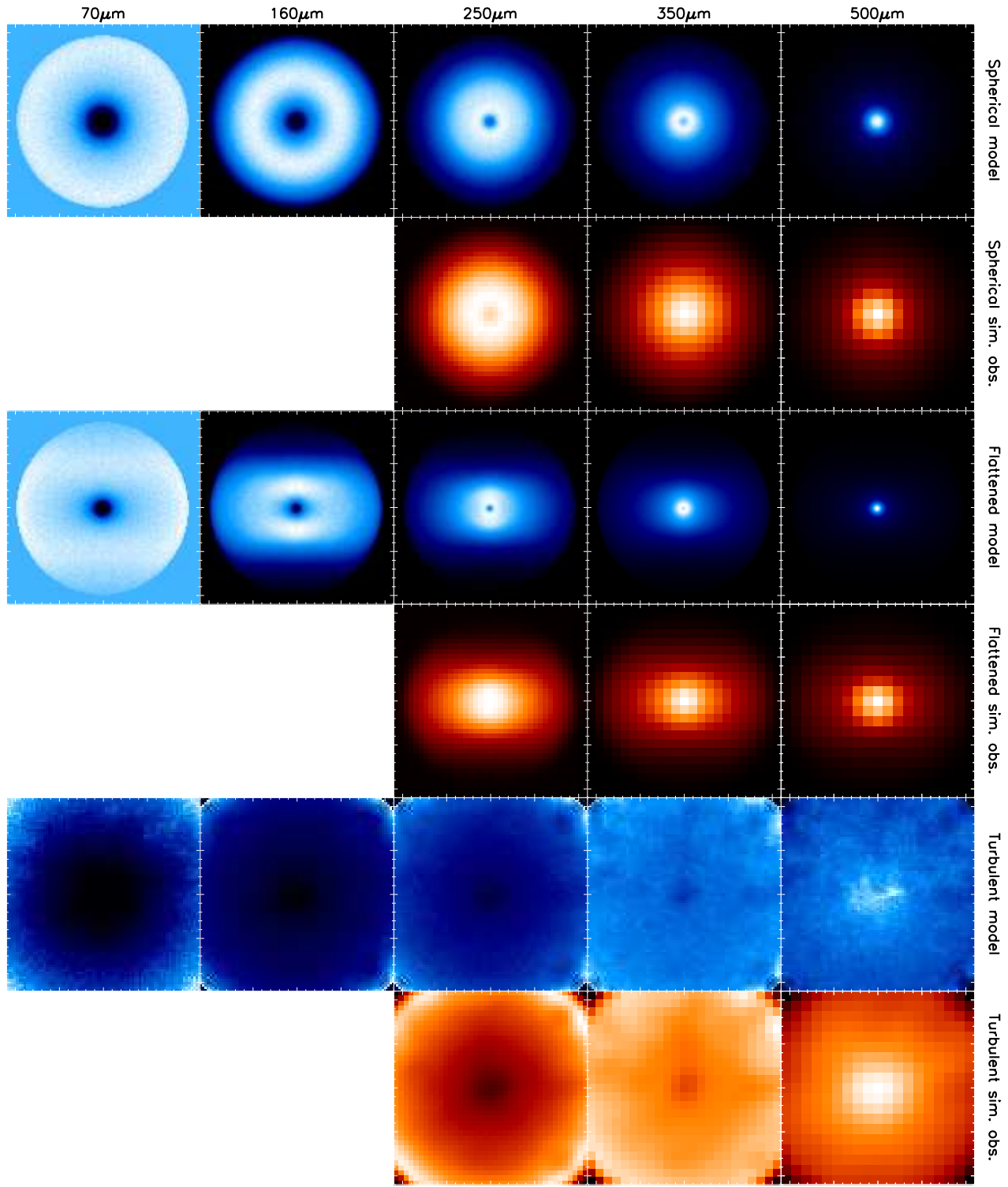}
	\caption{A grid of maps of the three model IRDCs shown by wavelength. The wavelength is listed at the top of each column. The first, third, and fifth rows show the intensity maps from the radiative transfer modelling for the spherical, flattened, and turbulent clouds respectively. The second, fourth, and sixth rows show the map from the 
row above (250-500~$\mu$m only) after it has been processed by the SPIRE telescope simulator. The RT models are shown with a blue colour table, the simulated observations are shown with a red colour table. The individual maps are linearly scaled between zero intensity (black) and maximum on the map (white). }
	\label{fig8:maps}
\end{figure*}

Synthetic maps of the plane-of-sky intensity distribution of the three IRDCs are produced using the radiative transfer model. The maps are produced at eight wavebands (8, 15, 70, 160, 250, 350, 500, and 500~$\mu$m) that replicate the filter sets of recent telescope facilities. The 8 and 15~$\mu$m wavelengths match  two bands from the MSX satellite (Price et al. 2001). The near infra-red  8~$\mu$m is also found on {\it Spitzer}'s IRAC camera (Fazio et al. 2004). The 8~$\mu$m filter can show strong emission from PAH grains (Flagey et al. 2006). { The radiative transfer model does not include transiently heated small grains. These grains dominate the emission at some IR bands (e.g. Li \& Draine 2003). In this paper we focus at longer wavebands ($\lambda>70\mu$m), where there is no emission from such grains.}

The far-infrared bands at 70 and 160~$\mu$m bands match approximately the 70 and 160~$\mu$m bands of {\it Spitzer}'s MIPS camera (Rieke et al. 2004), the N60 and N160 bands of AKARI (Kawada et al. 2007), and the 70 and 160 bands of {\it Herschel}'s PACS\footnote{PACS has been developed by a consortium of institutes led by MPE (Germany) and including UVIE (Austria); KU Leuven, CSL, IMEC (Belgium); CEA, LAM (France); MPIA (Germany); INAF-IFSI/OAA/OAP/OAT, LENS, SISSA (Italy); IAC (Spain). This development has been supported by the funding agencies BMVIT (Austria), ESA-PRODEX (Belgium), CEA/CNES (France), DLR (Germany), ASI/INAF (Italy), and CICYT/MCYT (Spain).} camera (Poglitsch et al. 2009, 2010). The submillimetre bands at 250, 350, and 500~$\mu$m match   approximately the filter set of {\it Herschel}'s SPIRE (Griffin et al. 2009, 2010) camera and the BLAST balloon experiment (Pascale et al. 2008). The 850~$\mu$m band matches approximately the long-wavelength filter of SCUBA (Holland et al. 1999) and SCUBA2 (Holland et al. 2006) on the JCMT. The 500~$\mu$m band is also roughly equivalent to SCUBA's and SCUBA2's 450~$\mu$m band (Holland et al. 1999; 2006). 

Given a synthetic map of intensity it is possible to simulate the observing process and produce simulated observations that match the resolution and expected instrumental characteristic of the chosen camera. As described in Section~\ref{sec:spire}, we have used the SPIRE  simulator to produce simulated observations for the 250--500~$\mu$m data. 

\begin{figure*}
	\includegraphics[width=\textwidth]{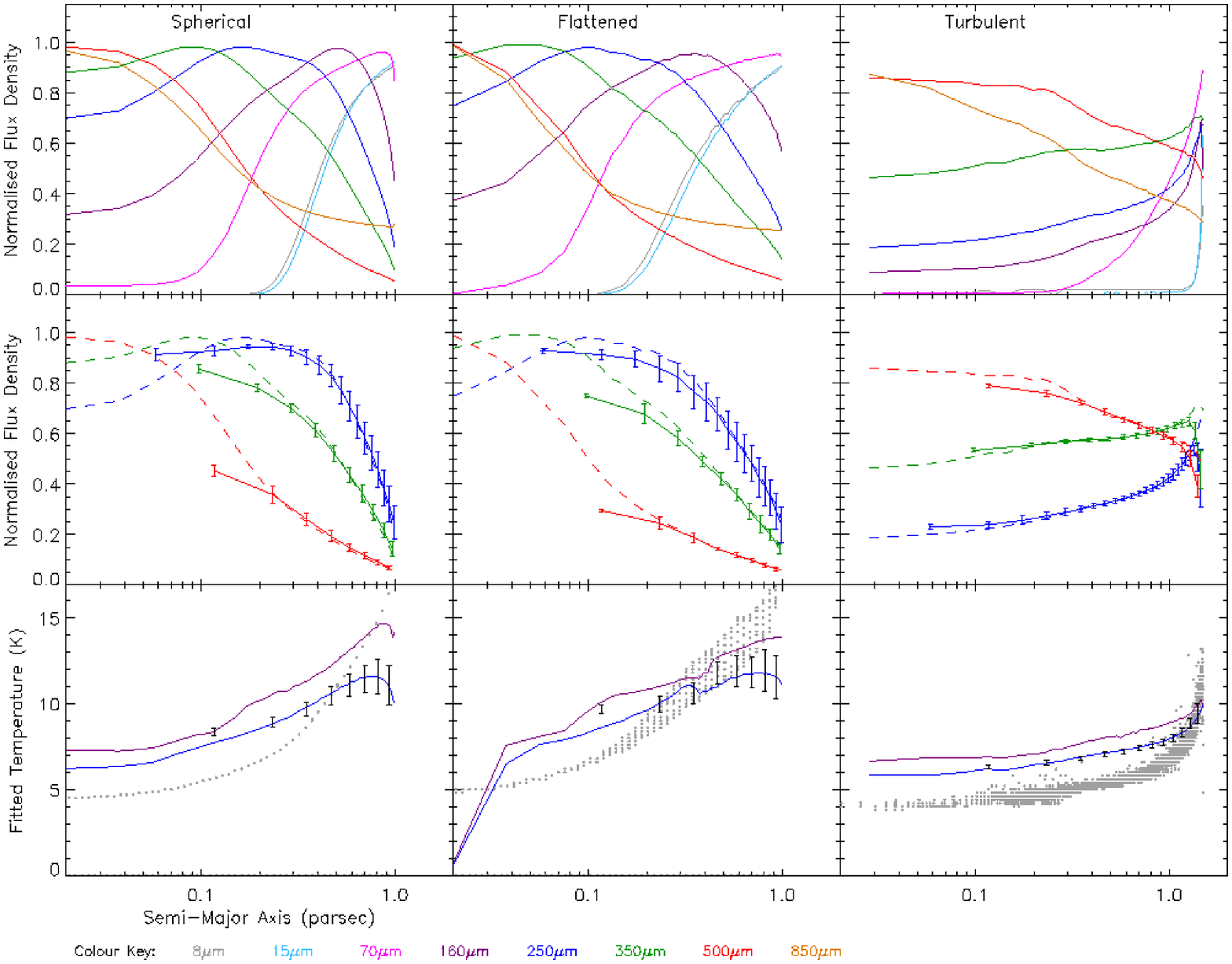}
	\caption{\label{fig9:prof} Radial profiles of the three model IRDCs at wavelengths $8-850$~\micron\ (top row).  Flux from the simulated observations (solid lines and error bars) versus the RT model flux (dashed lines) at the SPIRE bands (middle row). SED fitted temperature using just the SPIRE bands (blue curve) and the SPIRE bands plus 160~$\mu$m (purple curve). The equivalent fit using the three SPIRE simulated maps is shown by the error bars. The dots correspond to the actual dust temperature computed by the radiative transfer models (bottom row).}
\end{figure*}

Figure~8 shows the model IRDCs before (blue colour table) and after (red colour table) they have been processed into simulated observations. The columns are labelled at the top with the wavelength; the rows are labelled on the right-hand side with the description of the cloud. The maps  show a radius of 1.1 parsec around the centre of each cloud. The pixel size of the model cloud maps at all wavelengths is 0.019 pc (1.9 arcsec) for the spherical and flattened cloud, and 0.028 (2.9 arcsec) for the turbulent cloud. The pixel sizes of the 250, 350, and 500~$\mu$m maps are approximately a third of the SPIRE beam FWHMs, corresponding to 0.058 pc (6 arcsec), 0.097 pc (10 arcsec), and 0.116 pc (12 arcsec) at 250, 350, and 500~$\mu$m respectively. The 8- and 15-$\mu$m bands are not shown in Figure~8, as they simply show up  as dark clouds against a bright background.

IRDCs have been identified in large numbers from infra-red surveys (Simon et al. 2006; Peretto \& Fuller 2009) as higher-mass and more distant  analogues to the optically dark clouds of gas and dust seen in silhouette against a bright background in V-band surveys (e.g. Barnard 1927; Lynds 1962). Starting at the shortest wavelength in Figure~8 (70~$\mu$m) this extinction can be seen in the spherical and flattened model clouds as a central dark patch. The size of this patch decreases with increasing wavelength until 500~$\mu$m where it has either closed up entirely or can no longer be resolved. We note that the size of the emitting region at 500~$\mu$m is approximately  the same size as the dark region at 70~$\mu$m.

The substructure of the turbulent cloud which is evident in the model images is hardly visible in the simulated images due to the angular resolution of the simulations being larger than the length scale of the substructure at the assumed distance of the IRDC (2~kpc).

Table \ref{tab:fwhm} lists the FWHM of 2D Gaussians fitted to each of the 250--500~$\mu$m model clouds. These show quantitatively that the FWHM of the emitting region also decreases with increasing wavelength. This can be ascribed to the centre-to-edge temperature gradient. Cold dust located towards the centre of the cloud will preferentially radiate at longer wavelengths, whereas warmer dust located towards the edge of the cloud will preferentially radiate at shorter wavelengths. The wavelength 
dependent size of the cloud is also visible in the simulated observations. However, the central extinction hole is only resolved for the spherical cloud at 250~$\mu$m.

\begin{table}
\caption{The geometry of the plane-of-the-sky intensity distribution 
towards each of the model clouds.}
\label{tab:fwhm}
\begin{tabular}{lccc}
\hline
IRDC	 & \multicolumn{3}{c}{Observed FWHM (pc)} \\
	 & 250$\mu$m & 350~$\mu$m & 500~$\mu$m \\
\hline
Spherical & 0.563 $\times$ 0.565 & 0.474 $\times$ 0.474 & 0.262 $\times$ 
0.263 \\ 
Flattened & 0.565 $\times$ 0.368 & 0.509 $\times$ 0.331 & 0.341 $\times$ 
0.218 \\
Turbulent & --- & --- & 1.31 $\times$ 1.31 \\
\hline
\end{tabular}
\end{table}

\subsection{Radial Profiles}

Figure~9 shows radial profiles of each of the IRDCs shown in Figure~8. Each column shows a single model cloud as annotated in the upper plot (spherical, flattened, or turbulent) while each row of panels represents a particular property or set of parameters. No a priori knowledge (e.g. about the density profile and the dust temperature) is assumed whilst generating the radial profiles.

We compute an equivalent semi-major axis $a_{i}$ for each pixel $i$ with coordinates $(x_i,y_i)$ by assuming that it lies on an ellipse with the same origin, position angle, and axis ratio $r_{350}$ as the equivalent 
350~$\mu$m cloud (i.e., the simulated observations use the axis ratio of the 350~$\mu$m simulated cloud). The spherical and turbulent models have axis ratios of $\sim1.0$ and the flattened model cloud has an axis ratio of 
$\sim1.5$; it is assumed that the clouds' axes have a position angle of zero. The equivalent semi-major axis $a_{i}$ is given by 
\begin{equation}
a_{i}^{2} = (x_{\rm cen} - x_{i})^{2} + (r_{350}\times(y_{\rm cen} - y_{i}))^{2} 
\end{equation}
where $(x_{\rm cen}, y_{\rm cen})$ are the coordinates of the centre of the cloud. 
Once $a_{i}$ is known, we bin the pixels into bins of width $\Delta a$ and 
compute the mean and standard deviation of the intensities of the pixels 
in each bin. For the model clouds we use a bin width equal to the pixel size. 
This is 2.9 arcsec (0.029 pc). 

The top row of Figure~9 shows the distribution of mean intensity  versus equivalent semi-major axis for each of the eight model bands. The  flux density has been normalised for clarity, but no minimum flux has been subtracted (i.e. zero on the relative scale equates to zero on the absolute scale). The colour key for each band is shown beneath Figure~9. All three clouds show the characteristic IRDC extinction holes at PACS equivalent wavelengths (70 and 160~$\mu$m). The spherical and flattened clouds are completely extinguished at 0.2--0.3 pc, but the turbulent cloud is extinguished across its entire radius (due to its higher mass). 

The middle row of Figure~9 shows the profiles of the three SPIRE bands. The model profile, from the upper row, is reproduced as a dashed line. Plotted over this is the profile of the simulated observation. The process of observing the clouds causes flux to be averaged out due to the resolution limit of the telescope. This moves flux towards regions of lower flux, i.e. outward or towards central holes, and is most pronounced for bands where there is a significant central dip in the model profiles.

The lower panel of Figure~9 shows the results of fitting an optically-thin $\beta=1.85$ (Ossenkopf \& Henning 1994) greybody to the radial spectral energy distribution of each cloud. The blue and purple curves show the best-fit temperature for a greybody fitted to the 250-500~$\mu$m and 160-500~$\mu$m model cloud profiles respectively (i.e. using the results of the radiative transfer model).  The error bars show the  greybody fit to the 250-500~$\mu$m simulated observation profiles (i.e. taking into account instrumental effects).  These are interpolated to the lowest common resolution (500~$\mu$m) before fitting the greybody. 

These figures show that temperatures fitted to the observed fluxes match the temperatures that would have been fitted to the model IRDCs in the absence of instrumental effects.  However the fitted temperatures are higher than  the actual temperatures { (i.e. the ones calculated with the radiative transfer model)} in the centre of the  cloud and lower than the actual temperatures at the outer parts of the cloud. This is because the dust temperature from the simulated observations corresponds to an averaged temperature in the observed column, just as in  real observations. 

{This demonstrates that for accurately determining  the dust temperatures in  IRDCs, detailed radiative transfer modelling is needed  along with multi-wavelength, spatially resolved observations. Accurate dust temperatures are important for determining the masses of these clouds. Previous studies have shown that overestimating cloud dust temperatures even by  a few degrees may lead to underestimating cloud masses by a factor of $2-3$ (Stamatellos et al. 2007). This has important implications for the inferred cloud stability and for the derived mass function of IRDCs.

The methodology to determine the actual properties of an IRDC  is to fit the observed fluxes at all available wavebands, by varying the assumed density structure of the radiative transfer model. The density structure can be spherically symmetric or axisymmetric (e.g. flattened core), depending on each specific cloud. 3-dimensional modelling is also possible but the determined density structure cannot uniquely recovered (e.g. Steinacker et al. 2003, 2005). Once the observed fluxes are fitted, the density profile of the cloud is determined, and the actual dust temperatures are calculated through the radiative transfer model. 
}

\section{G29.55+00.18, an IRDC from the Hi-GAL survey}

\begin{figure*}
\centerline{
\includegraphics[width=6.cm]{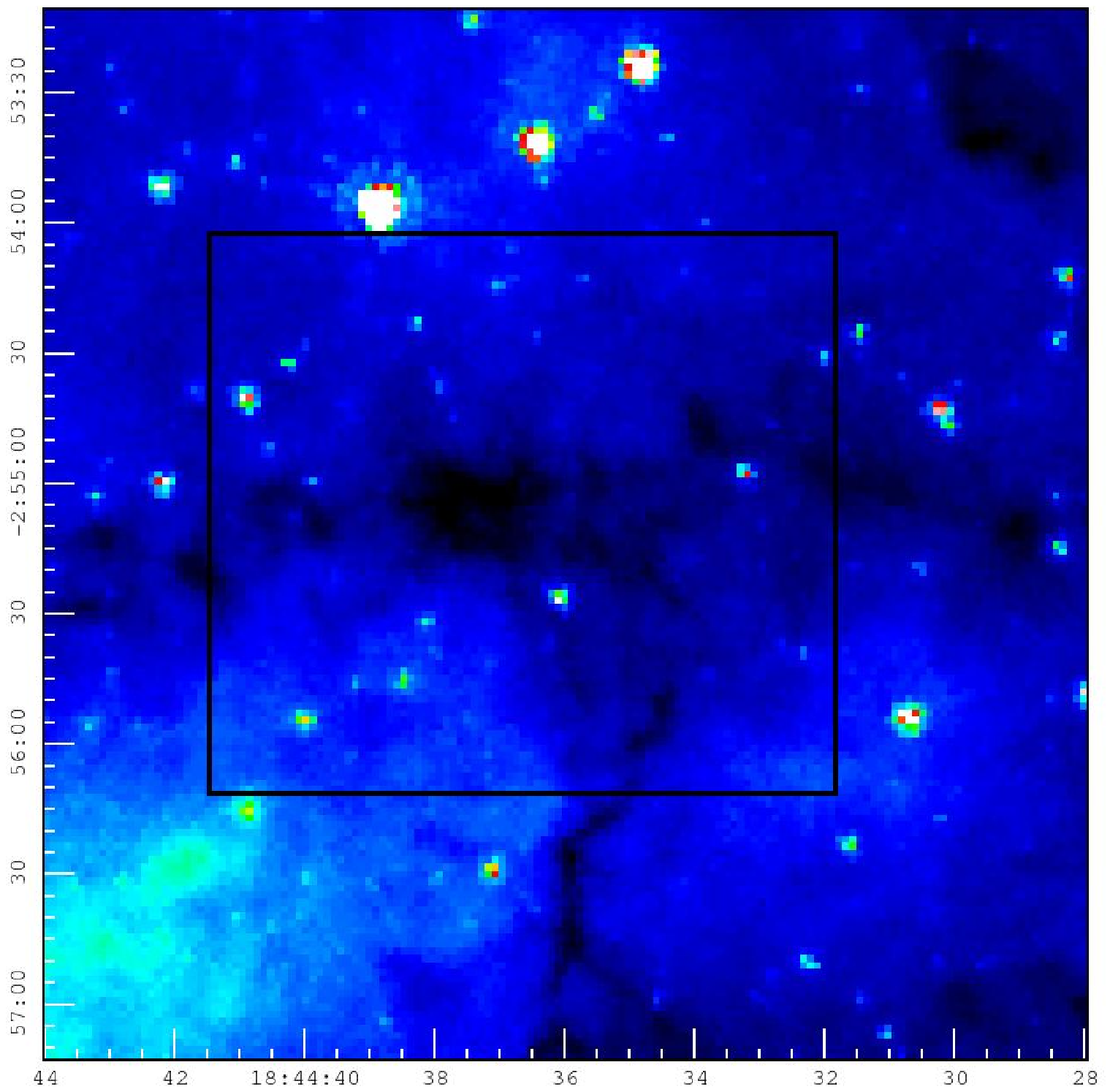}
\includegraphics[width=6.cm]{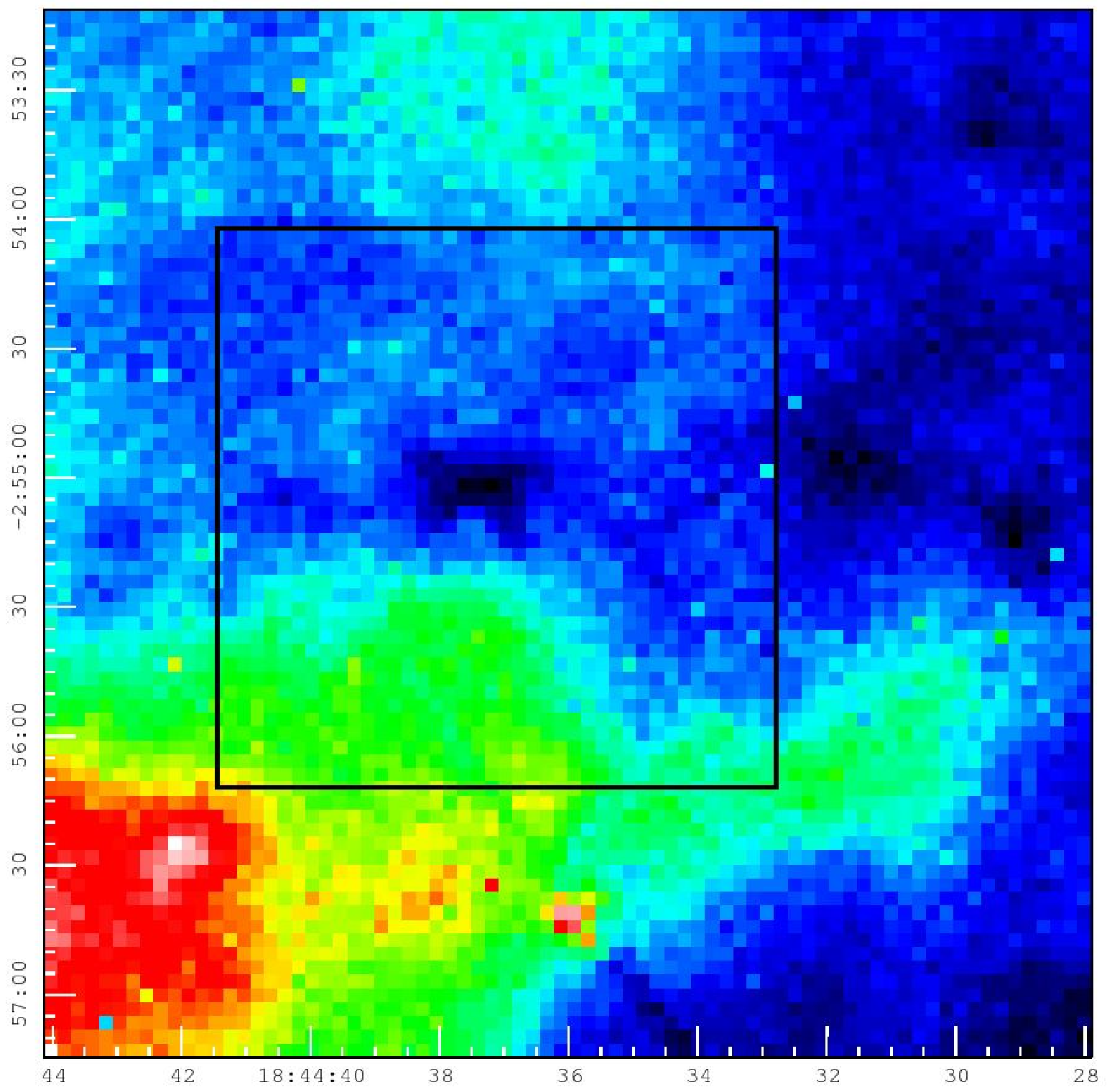}
\includegraphics*[viewport=0.3cm 0.3cm 19.cm 15.5cm, width=6.cm]{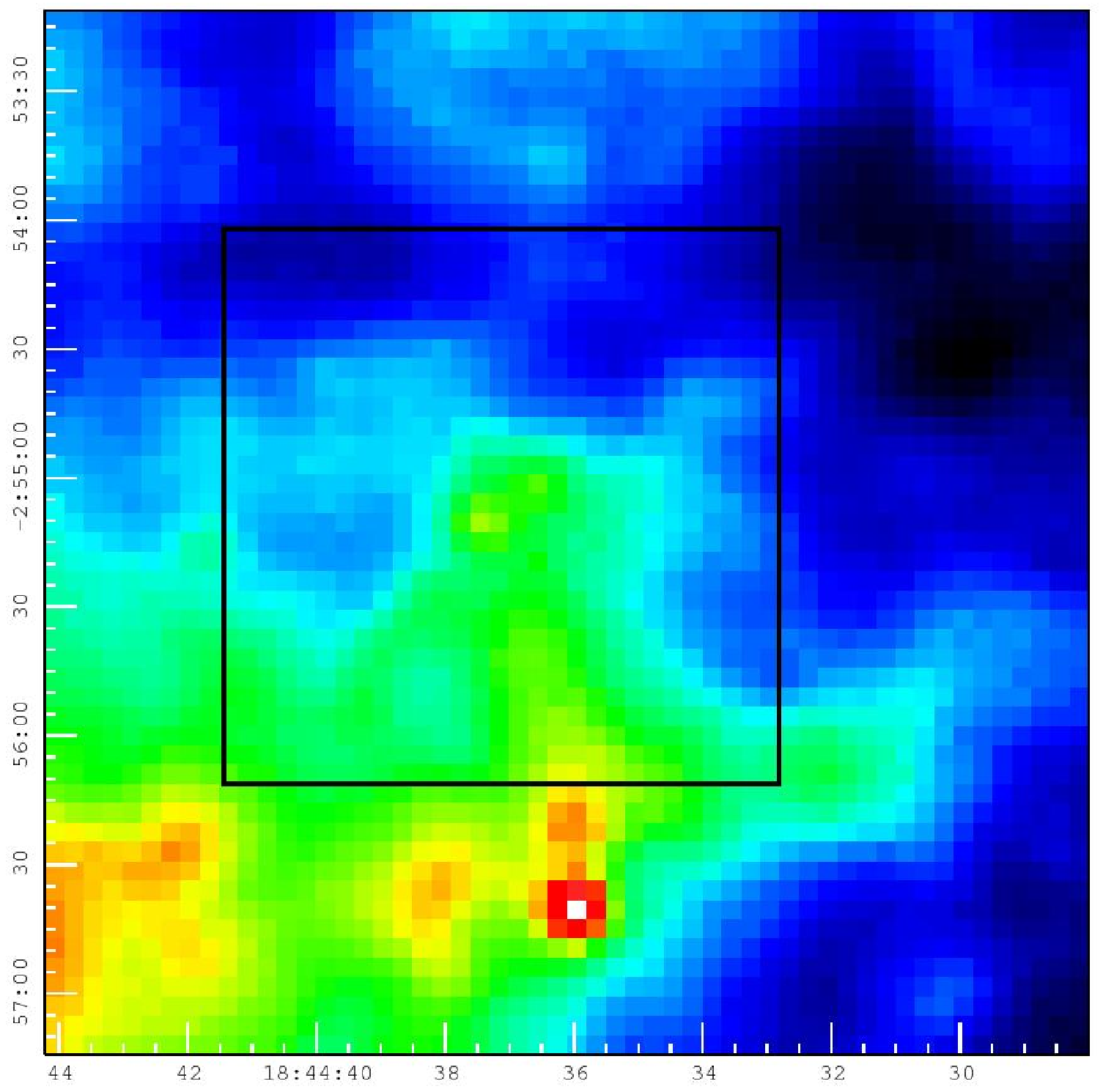}
}
\centerline{
\includegraphics[width=6.cm]{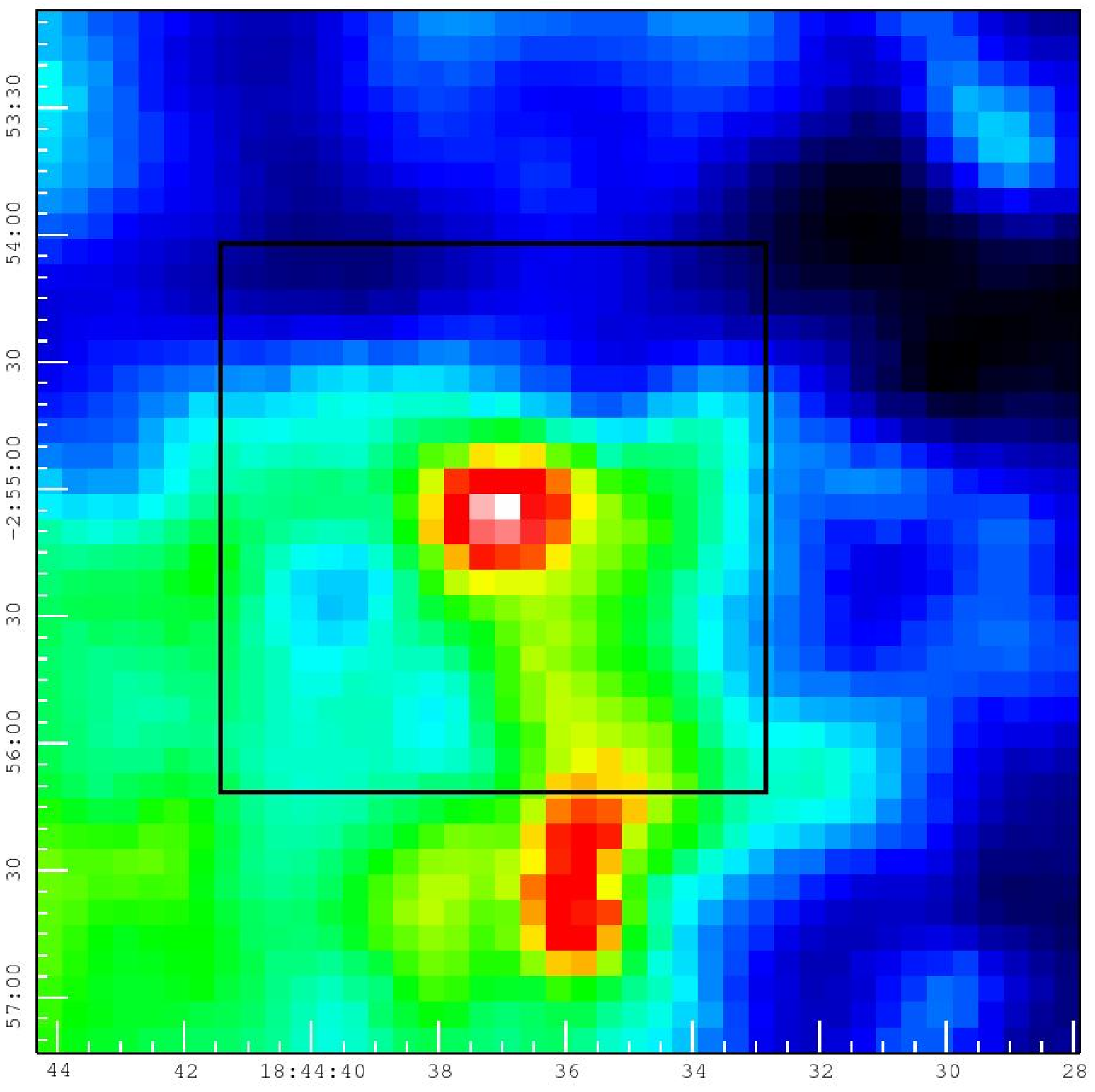}
\includegraphics*[viewport=0.3cm 0.3cm 19cm 15.5cm, width=6.cm]{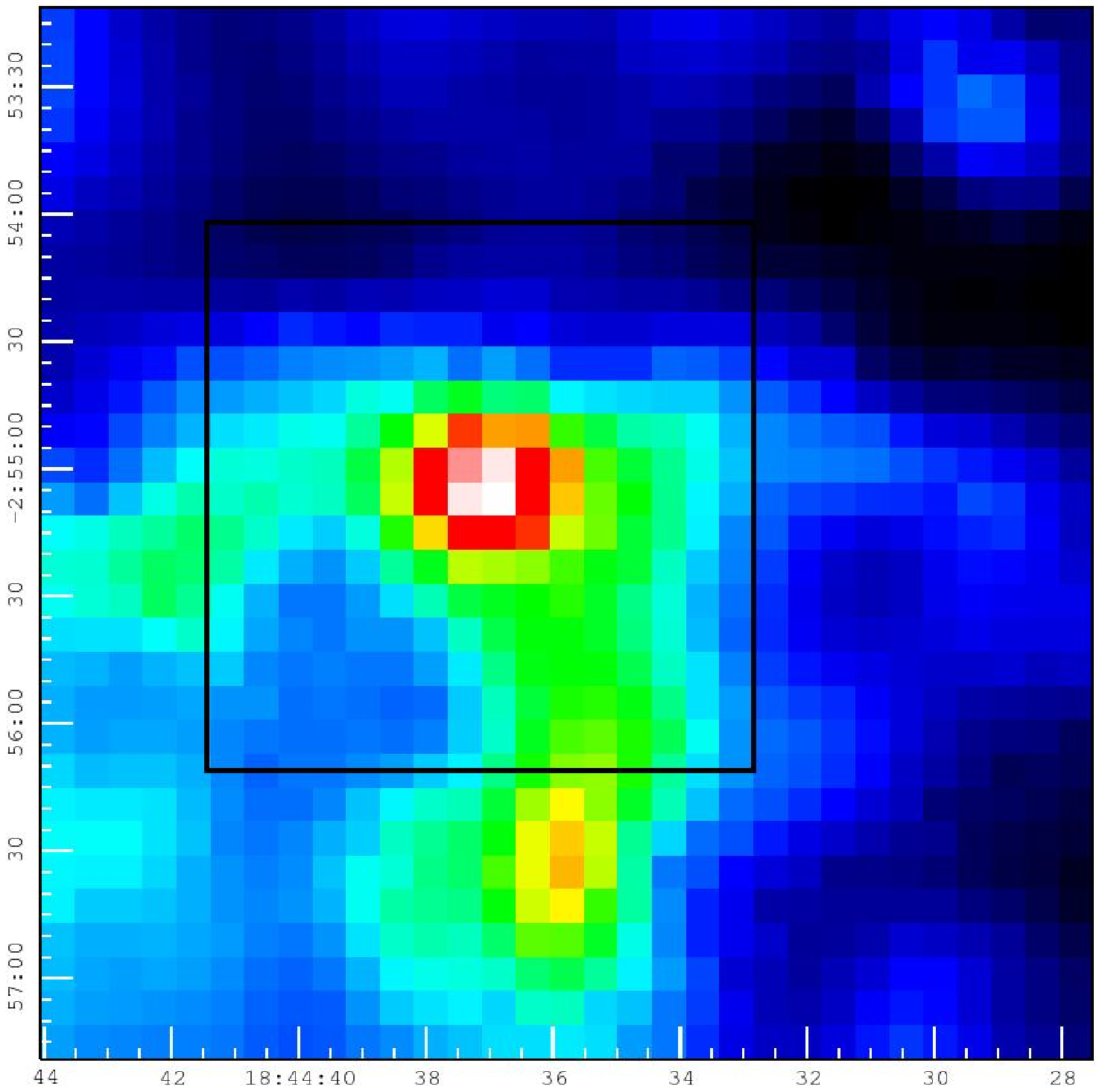}
\includegraphics*[viewport=2.5cm 2.2cm 16.5cm 14.cm, width=6.cm]{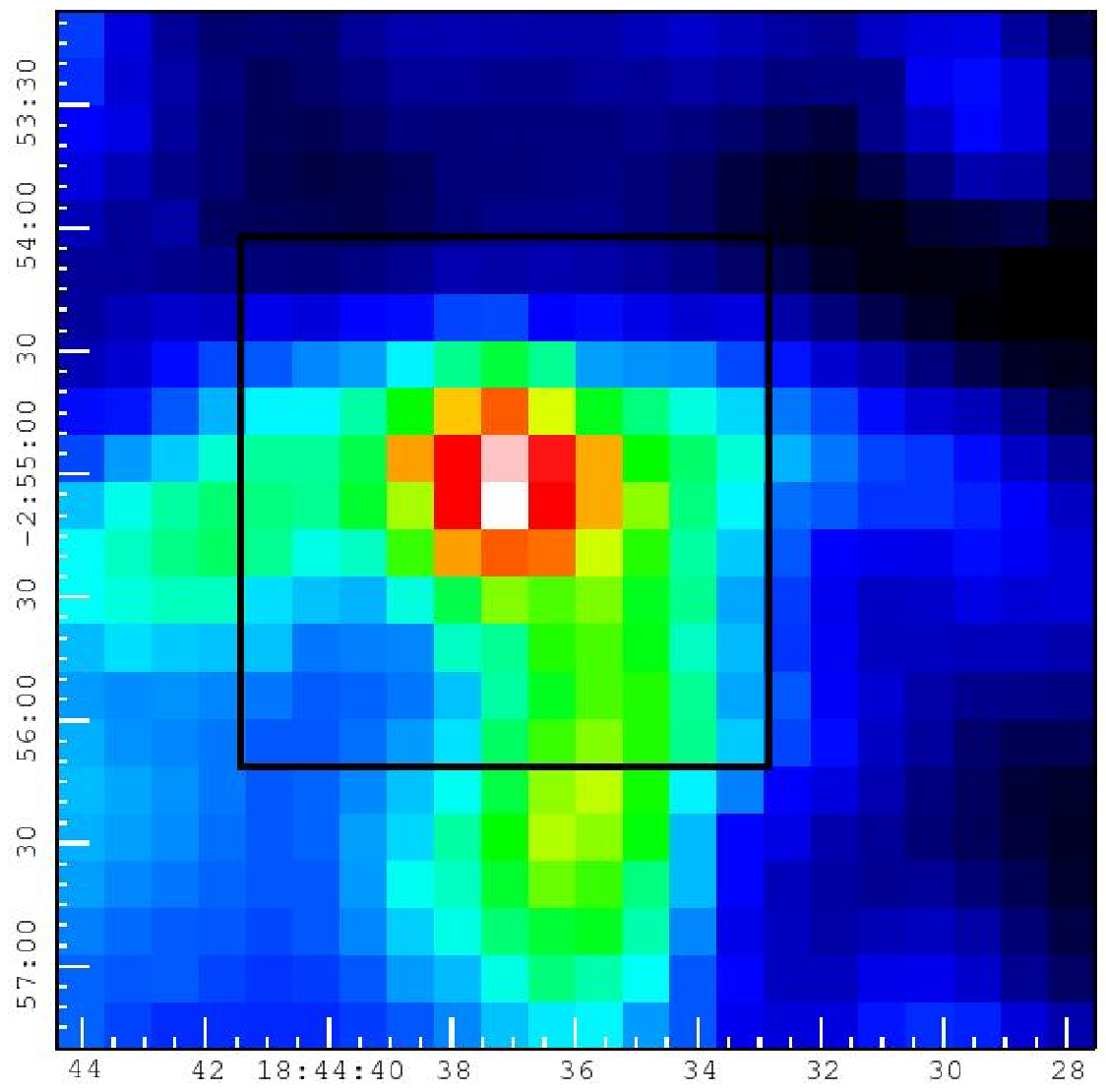}
}
\caption{G29.55+00.18 observations at 6 wavebands (top row: 8~$\micron$ {\it Spitzer}/GLIMPSE, 70~$\micron$ and 170~$\micron$  PACS; bottom row: 250~$\micron$, 350~$\micron$  and 500~$\micron$  SPIRE). The cloud is seen in absorption against the bright background at 8~$\micron$ and 70~$\micron$, and in emission at longer wavelengths.}
\label{fig:2918obs}
\centerline{
\includegraphics[width=5.5cm]{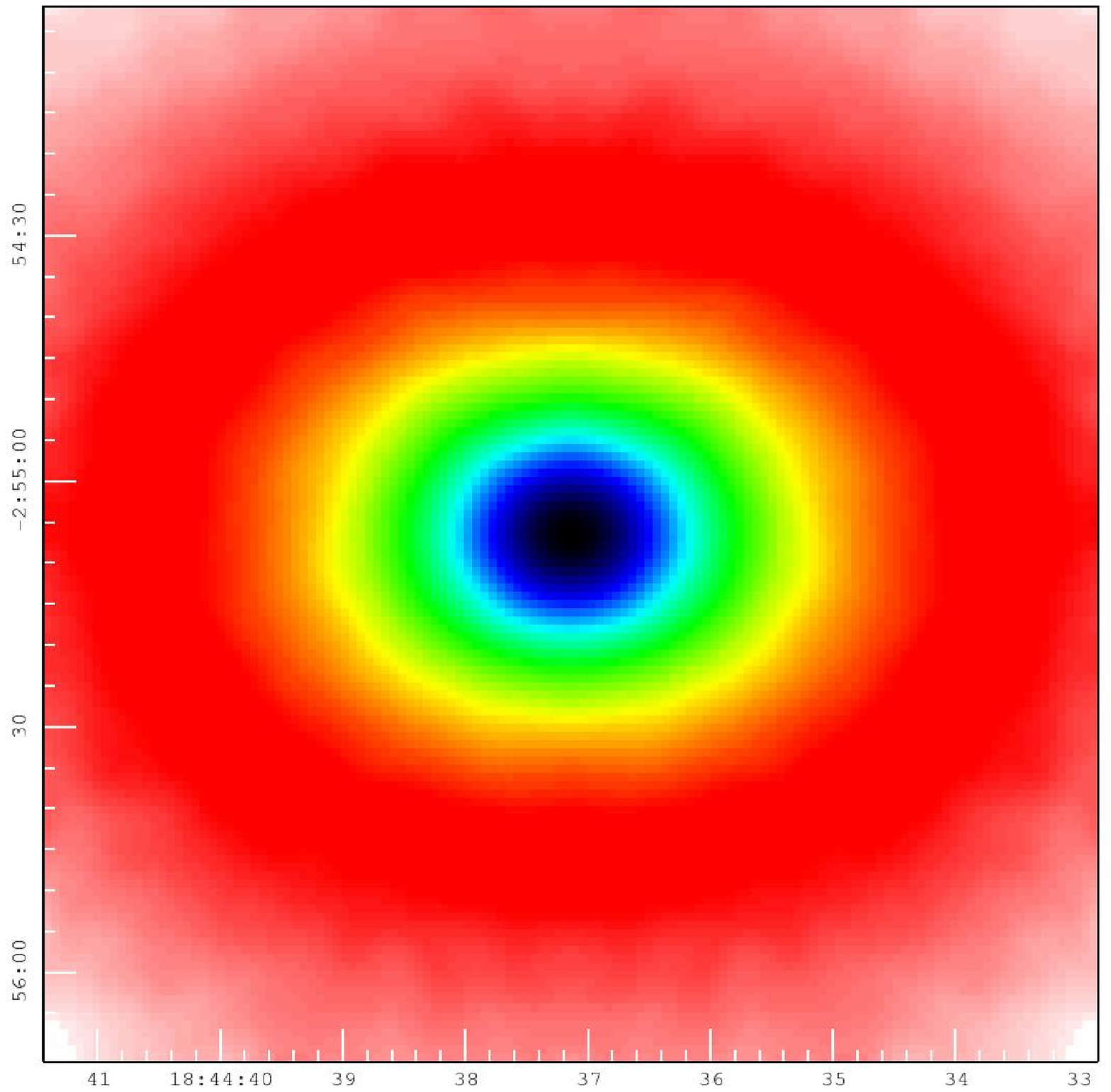}
\hspace{0.5cm}
\includegraphics[width=5.5cm]{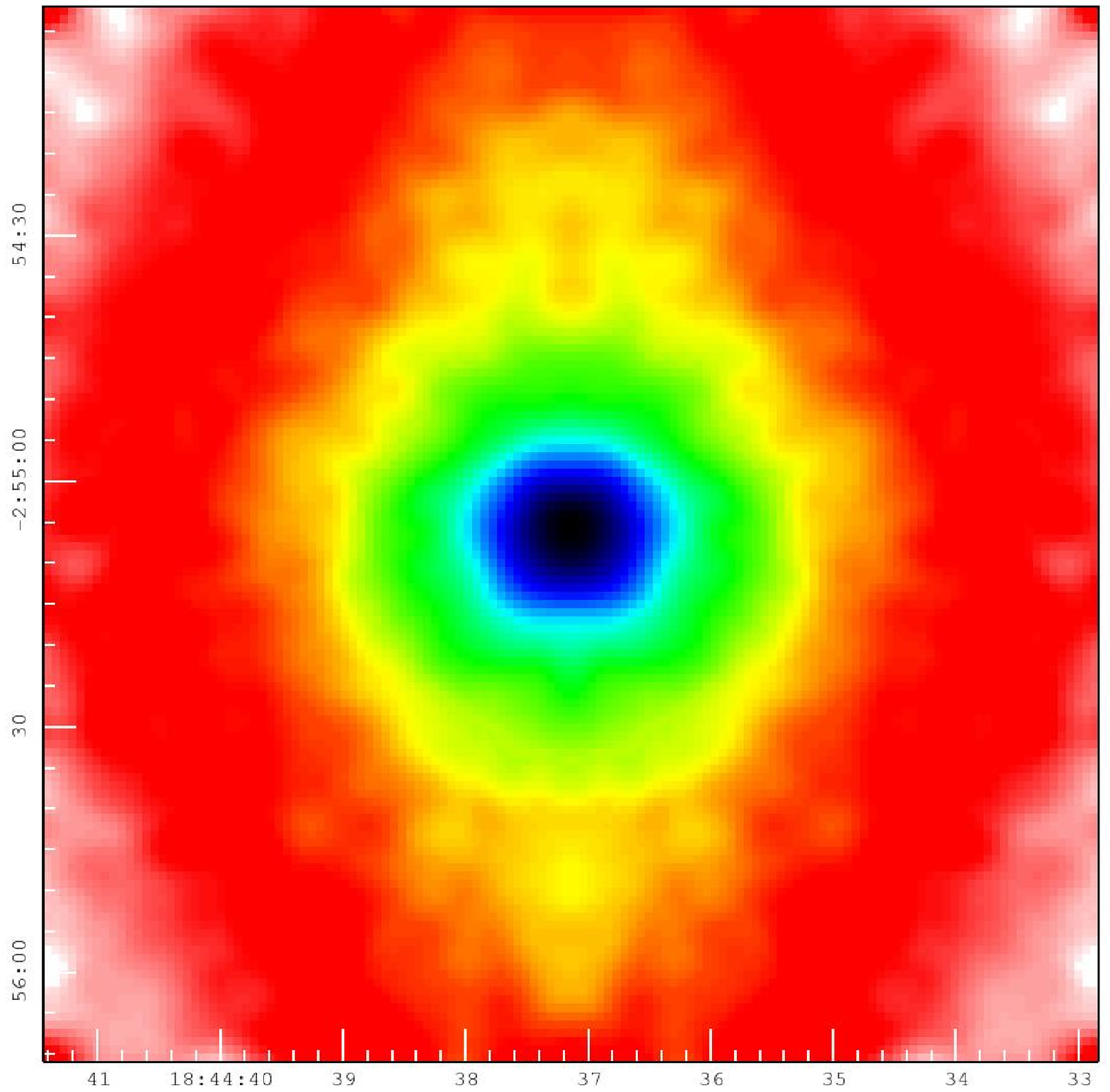}
\hspace{0.5cm}
\includegraphics[width=5.5cm]{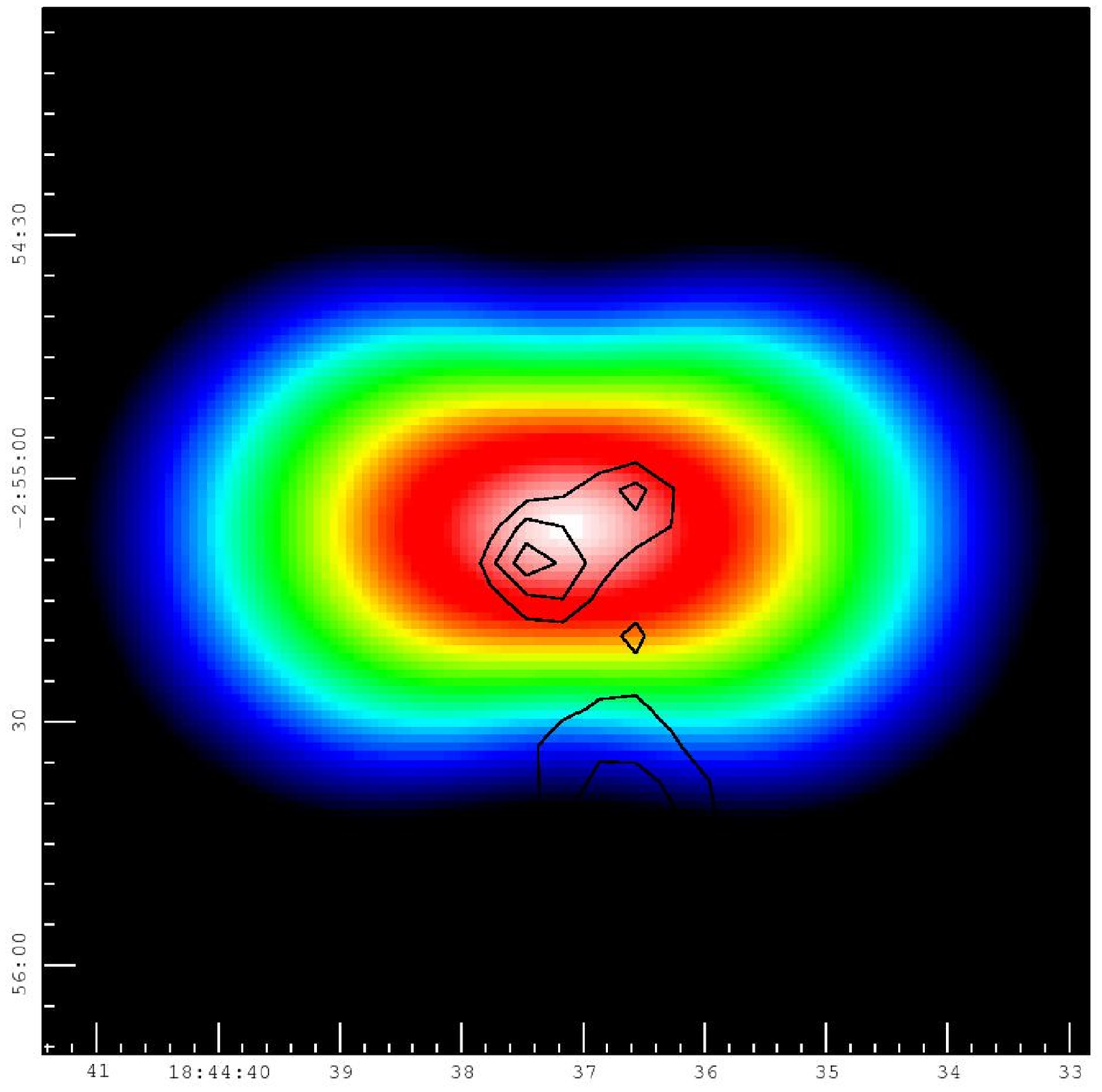}
}
\vspace{0.4cm}
\centerline{
\includegraphics[width=5.5cm]{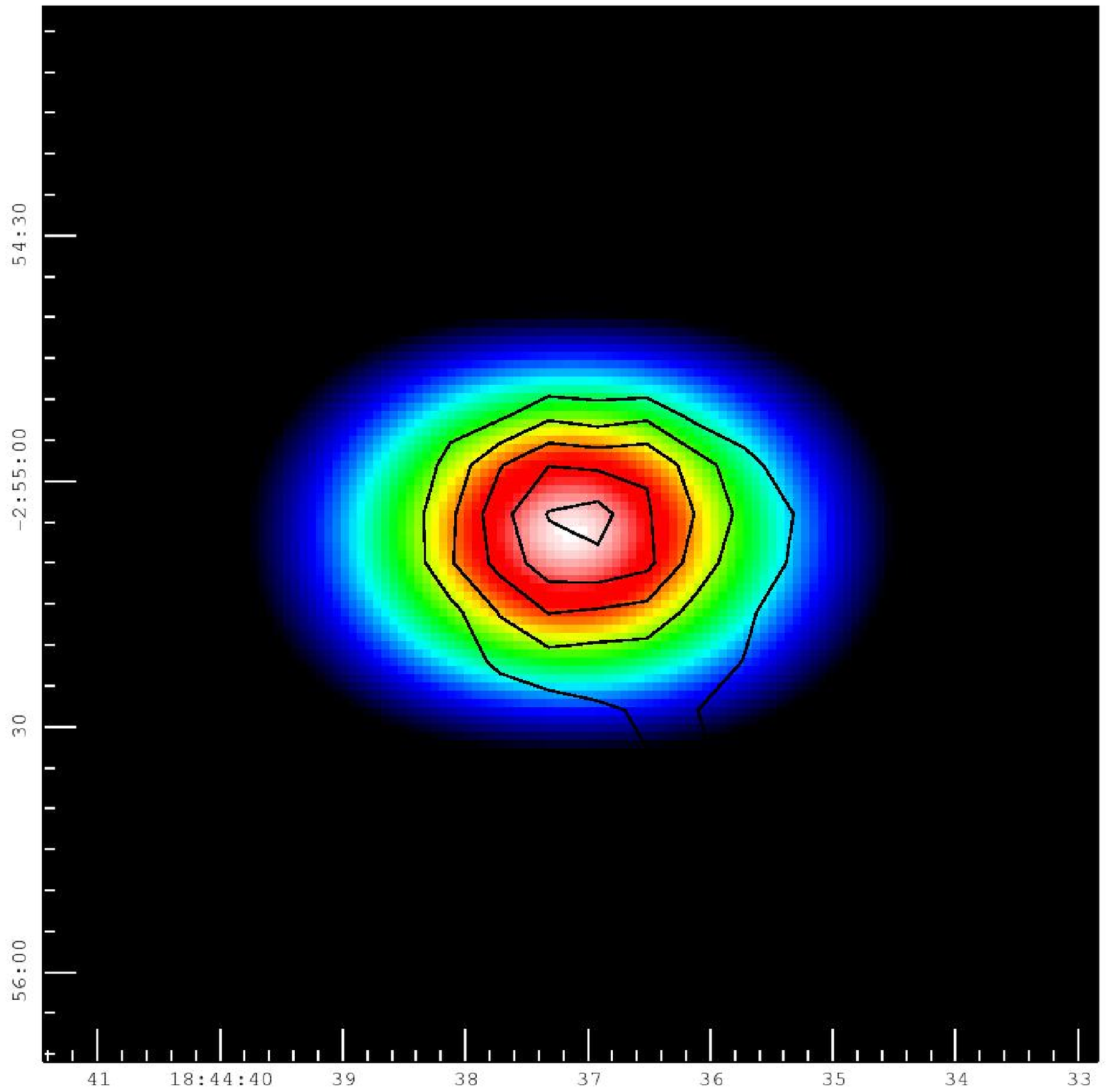}
\hspace{0.5cm}
\includegraphics[width=5.5cm]{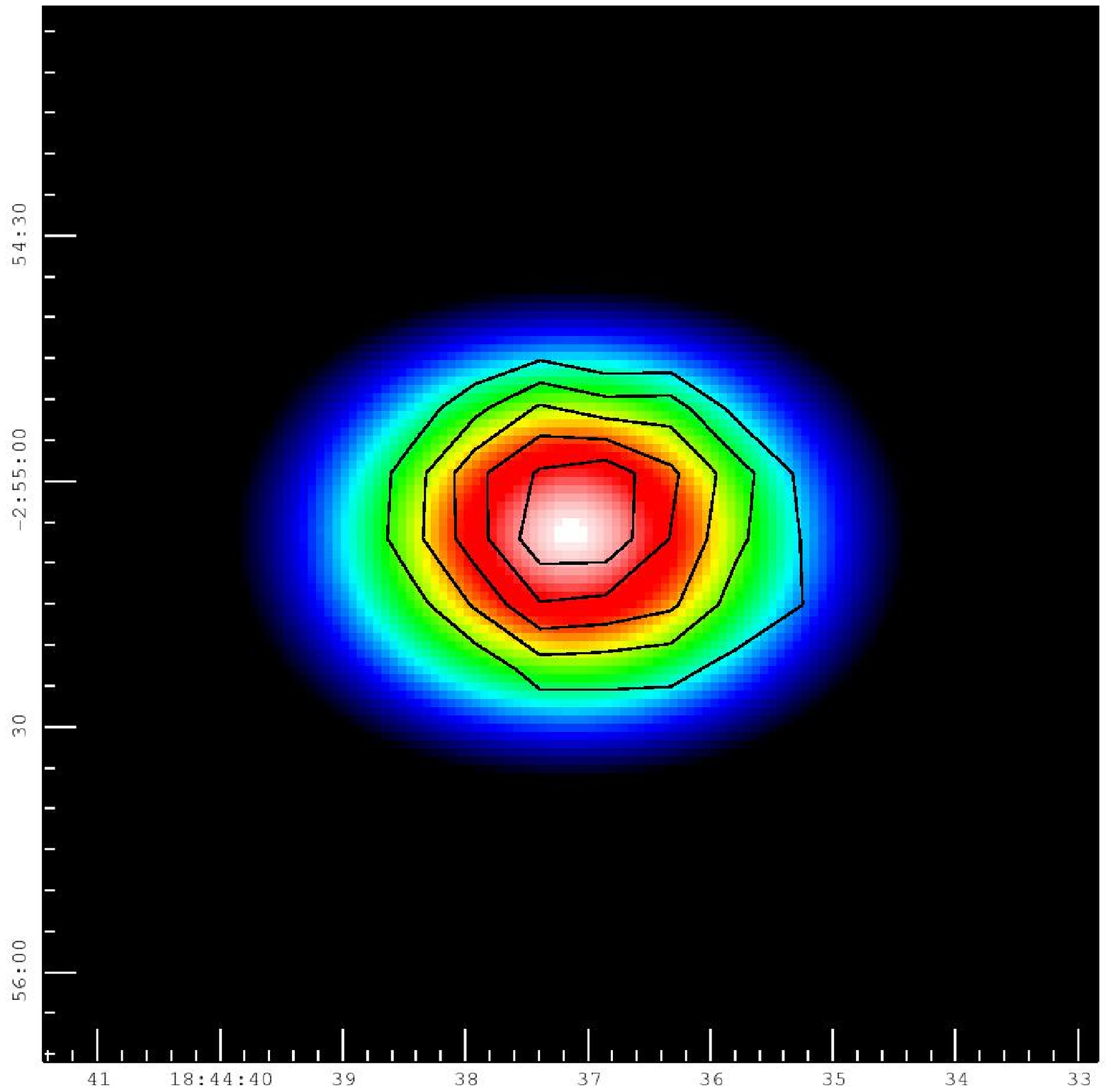}
\hspace{0.5cm}
\includegraphics[width=5.5cm]{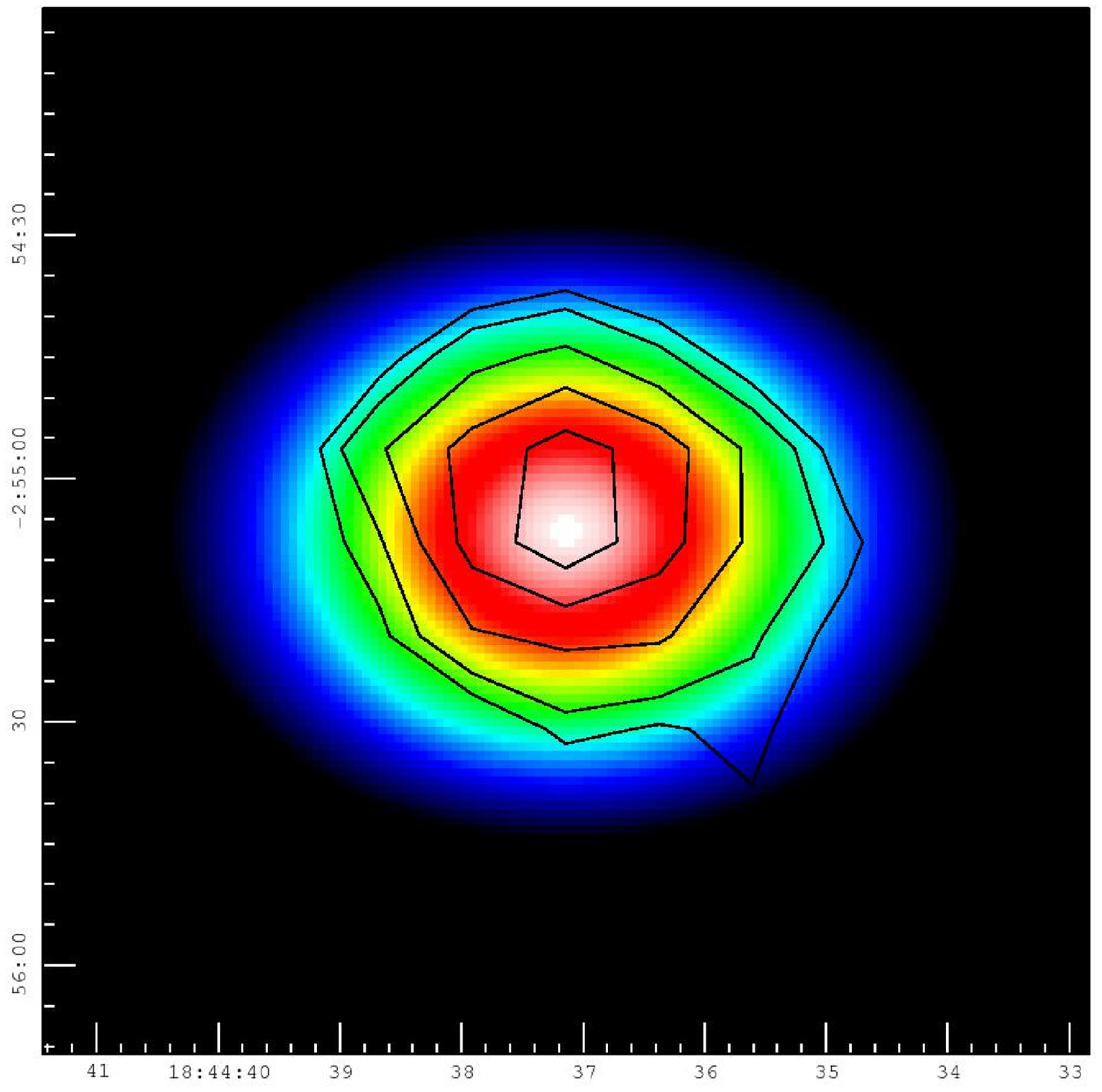}
}
\caption{G29.55+00.18 simulated observations at 6 wavebands (same as in Fig.~\ref{fig:2918obs}). Contours are taken from the observed data at the corresponding wavelength. The model reproduces the general characteristics of the appearance  the core (flux and shape).}
\label{fig:2918model}
\end{figure*}

\begin{figure}
\centerline{
\includegraphics[width=8.5cm]{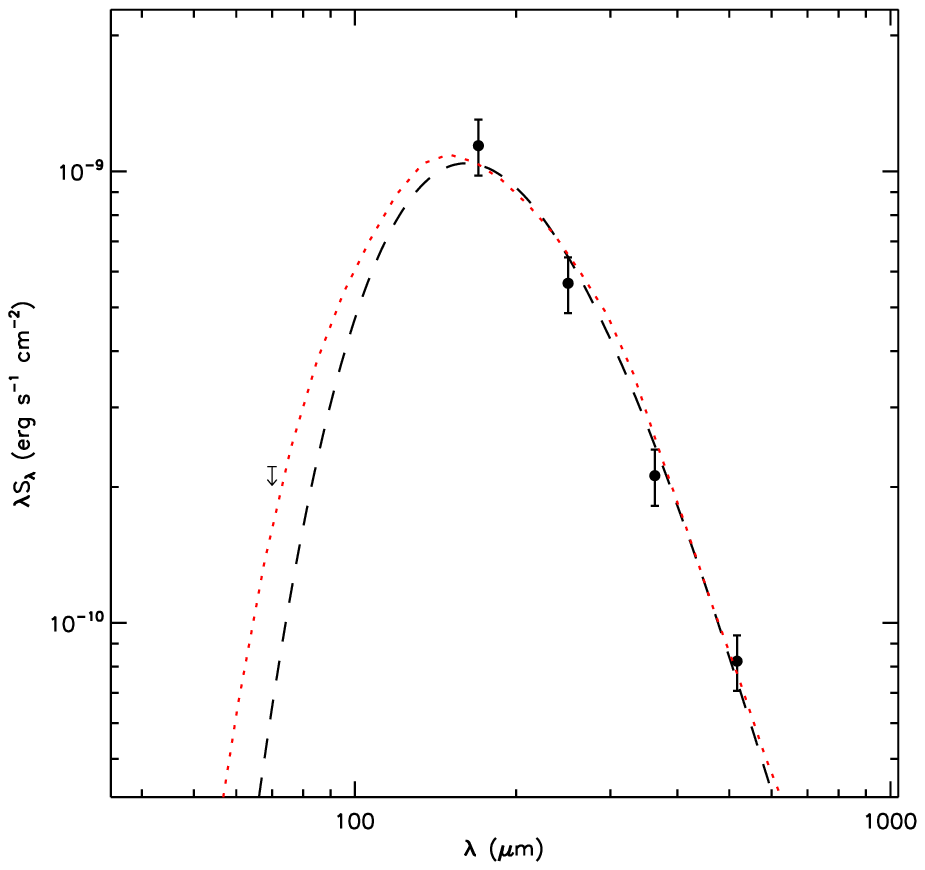}}
\caption{SED of G29.55+00.18. The dotted line corresponds to the SED calculated from the radiative transfer model, and the dashed line corresponds to the SED  from a single-temperature, grey-body fit.}
\label{fig:2918sed}
\centerline{
\includegraphics[height=8.cm,angle=-90]{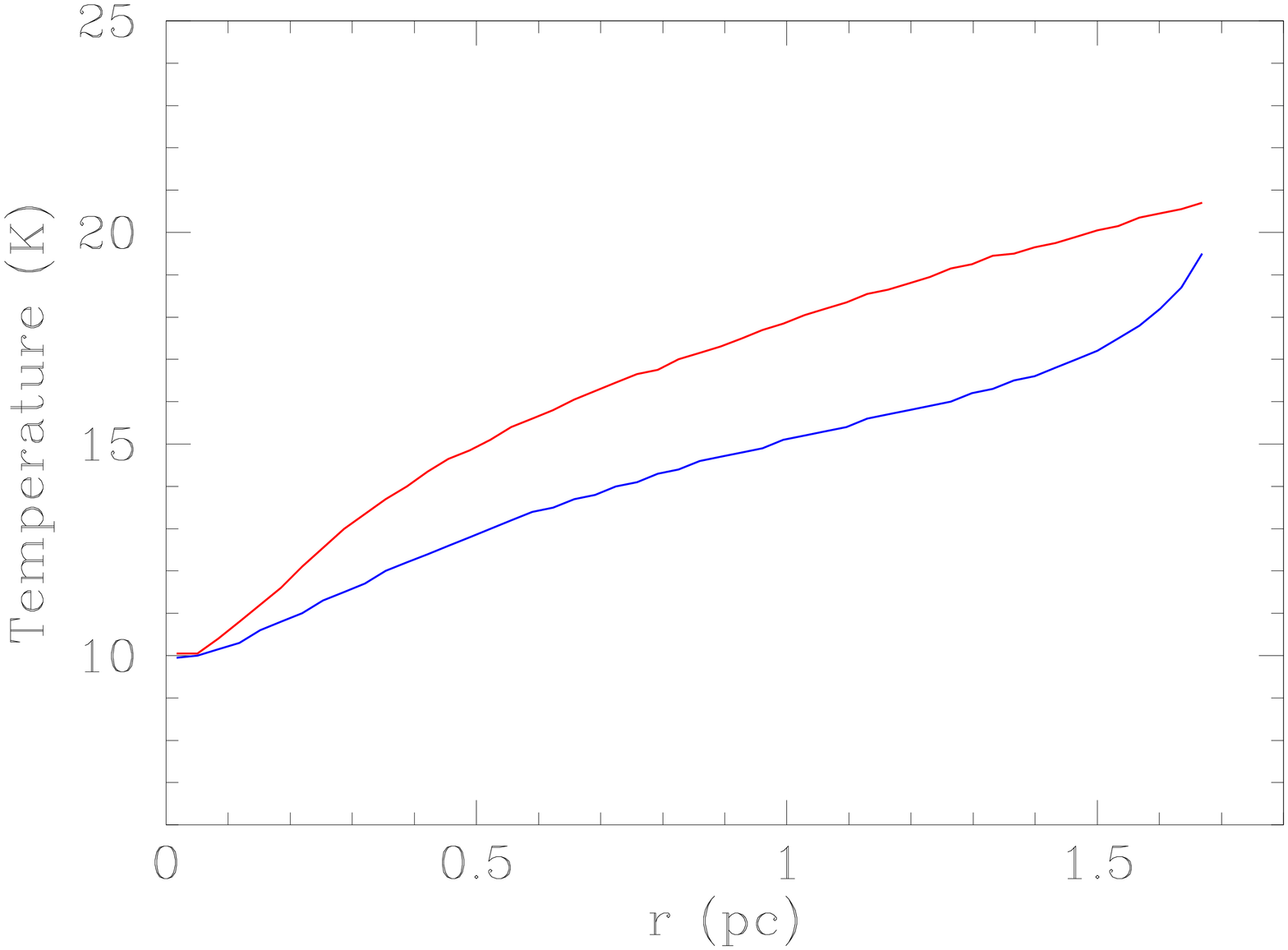}}
\caption{Dust temperature profile of  G29.55+00.18 at two different directions  in the cloud. The bottom line corresponds to the midplane of the flattened structure ($\theta=90\degr$), whereas the upper line corresponds to the direction perpendicular to the midplane ($\theta=0\degr$).}
\label{fig:2918temp}
\end{figure}

G29.55+00.18 (Simon et al. 2006) is an infrared-dark cloud, which has been observed by {\it Herschel}/PACS and {\it Herschel}/SPIRE as part of  the Hi-GAL survey, and by {\it Spitzer}/GLIMPSE.  The cloud is seen in absorption against the bright background at 8~$\micron$ and 70~$\micron$, and in emission at longer wavelengths (Fig.~\ref{fig:2918obs}). The SED of the cloud is shown in Fig.~\ref{fig:2918sed}.

\begin{table}
\centering
\caption{The physical properties of the infrared-dark cloud G29.55+00.18.  $R_{\rm cloud}$: radius, $r_0$ : flattening radius, $n_c({\rm H}_2)$ : central density, 
$\tau_{\theta=90\degr}/\tau_{\theta=0\degr}$: ratio of the visual optical depths,
$f_{\rm ISRF}$: ISRF factor (see text), 
$T_{\rm SED}$: temperature calculated from a single-temperature, grey-body fitting of the SED, $T_{\rm model}$: temperature from the radiative transfer model,
$M_{\rm SED}$: mass calculated from Hildebrand (1983) and $T_{\rm SED}$,
$M_{\rm model}$: mass calculated from the radiative transfer model.
The distance is taken from Heyer et al. (2009).}
\label{tab:2819}
\renewcommand{\footnoterule}{}  \centering   
\begin{tabular}{lc}
\multicolumn{2}{c}{G29.55+00.18}\\
\hline 
R.A. (2000)						&18:44:37.00\\
Dec. (2000)						&-02:55:07\\
Distance (kpc)						&4.8\\
$R_{\rm cloud}$ (pc)							&1.7\\
$r_0$ (pc)						&0.2\\
$n_c({\rm H}_2)$ (cm$^{-3}$)			& $2.8\times10^{4}$\\
$\tau_{\theta=90\degr}/\tau_{\theta=0\degr}$ & 1.55\\	
$f_{\rm ISRF}$						&2.5\\
$T_{\rm SED}$ (K) 					& 16\\
$T_{\rm model}$ (K)					& $10-21$\\
$M_{\rm SED}$ (M$_{\sun}$)			& 530\\
$M_{\rm model}$ (M$_{\sun}$)			& 520\\
\hline
\end{tabular}
\end{table}

Guided by the image of the cloud at 500~$\micron$, we  model this cloud using a 2D density profile (Eq.~\ref{eq:flatdens}) with a ratio of optical depths $\tau_{\theta=90\degr}/\tau_{\theta=0\degr}=1.55	$. The input parameters of the model and the derived physical properties of the cloud are listed in Table~\ref{tab:2819}.  The mass of the cloud is constrained by the longer wavelength data, which are approximately column density profiles, as the effect of the temperature gradient is relatively weak. The temperature of the cloud is mainly constrained by the shorter wavelength data. To match these short wavelength data, the external radiation field is enhanced by a factor of $f_{\rm ISRF}=2.5$, when compared with the ISRF at the solar neighbourhood, which is consistent with the higher ambient radiation field in the Galactic plane. 

The model reproduces well the images of the cloud at different wavelengths (Fig.~\ref{fig:2918model}). At shorter wavelengths (8, 70~$\micron$) the cloud is seen in absorption (note that the background is not modelled, hence there is no actual correspondence with the observed background), and in emission at longer wavelengths ($\ga170~\micron$). The flux and the shape of the cloud are well matched (assuming a viewing angle of  $\theta_{\rm obs}=40\degr$). At 170~$\micron$ there is secondary peak that is not reproduced by the model.  This is probably due to asymmetric heating as this secondary peak does not appear at longer wavelengths (cf. Nutter et al.~2009); such heating is not included in the current model. 

The observed SED is well fitted  by the model  (Fig.~\ref{fig:2918sed}, dotted line). The dust temperature inside the cloud drops from to 21~K at its edge  to 10~K at its centre (Fig.~\ref{fig:2918temp}; cf. Peretto et al. 2010). Additionally, there is a variance of the temperature with the polar angle $\theta$, with the more dense region at the "equator" of the cloud being colder by up to $\sim 3$~K, as compared with the corresponding equal-radius region at the "pole" of the cloud. The mass of the cloud is calculated to be 530~M$_{\sun}$.

We also compare the radiative transfer model with a simple grey-body single-temperature fitting of the SED. The SED is well reproduced using a temperature of 16~K (Fig.~\ref{fig:2918sed}, dashed line). The mass of cloud calculated using the above temperature and the flux at 500~$\micron$, is 520~M$_{\sun}$, which is slightly lower but consistent with the mass obtained from the radiative transfer model, despite using an average, and not the actual, dust temperature. However, the inner dense regions of the cloud where star formation will occur are colder (by $\sim$5~K) than the estimated average temperature.

This example demonstrates the need for  radiative transfer modelling for determining the temperature structure of IRDCs. Despite the fact that a simple grey-body single-temperature model provides a good fit to the SED, this model cannot accurately determine the temperature structure in the cloud and overestimates the temperature of the interesting, in terms of potential for star formation, inner dense regions of the cloud.

\section{Conclusions}

We have demonstrated the use of the 3D, multi-wavelength Monte Carlo code {\sc phaethon}, to model the transfer of radiation in IRDCs, that are externally illuminated by the interstellar radiation field. The cores of these clouds are believed to be where high-mass stars form, i.e. they are the high-mass equivalent of prestellar clouds. We have presented three widely different models, in which we varied the mass, density, radius, morphology and internal velocity field of the cloud. We have shown the predicted output of the model at the wavebands of {\it Herschel} and {\it Spitzer}. We also passed the model output through the SPIRE simulator to produce simulated observations of these IRDCs. These were then analysed as if they were real observations. Subsequently, the results of this analysis were compared with the results of the radiative transfer modelling. 

Our study highlights the need for detailed radiative transfer modelling  when using multi-wavelength observations from {\it Herschel} to  accurately determine the properties of IRDCs.  This method was applied for the study of G29.55+00.18, an IRDC from the Hi-GAL survey. Modelling of a larger sample of IRDCs found in the Hi-GAL survey will appear in a future publication (Wilcock et al., in prep).

\section*{Acknowledgements}
We would like to thank the referee for his report that helped improving the original manuscript. Simulations were performed using the Cardiff HPC Cluster {\sc merlin}. The colour plots of Fig.~5 were produced using {\sc splash} (Price 2007). DS and JMK acknowledge post-doctoral support from the Science \& Technology Facilities Council (STFC) under the auspices of the Cardiff Astronomy Rolling Grant.

\end{document}